\documentclass[aps,pra,twocolumn,showpacs,floats]{revtex4}
\usepackage{latexsym}
\usepackage{amsmath}
\usepackage{amssymb}
\usepackage{bm}
\usepackage{wasysym}
\usepackage[dvips]{color}
\usepackage{graphicx}
\usepackage{graphicx}
\usepackage{subfigure}
\usepackage{epsfig}

\newcommand{\veps}{\varepsilon}

\newcommand{\ra}{\rangle}
\newcommand{\la}{\langle}
\newcommand{\ua}{\uparrow}
\newcommand{\da}{\downarrow}
\newcommand{\bfr}{{\bf r}}

\definecolor{purple}{rgb}{0.5, 0, 1}

\def\mbfE{\mathbf{E}}
\def\mbfe{\mathbf{e}}

\def\mbfB{\mathbf{B}}

\def\e{\mathrm{e}}

\def\mbfq{\mathbf{q}}

\def\mbfp{\mathbf{p}}

\def\mbfd{\mathbf{d}}
\def\mbfa{\mathbf{a}}

\def\mbfr{\mathbf{r}}

\def\mbfF{\mathbf{F}}

\def\3P0{$^{3}$P$_{0}$}
\def\1S0{$^{1}$S$_{0}$}

\usepackage{amsmath,stmaryrd,graphicx}
\usepackage{soul}

\makeatletter
\newcommand{\fixed@sra}{$\vrule height 2\fontdimen22\textfont2 width 0pt\shortrightarrow$}
\newcommand{\shortarrow}[1]{%
  \mathrel{\text{\rotatebox[origin=c]{\numexpr#1*45}{\fixed@sra}}}
}
\makeatother

\begin{document}

\title{Atoms trapped by a spin-dependent
  optical lattice potential: \\
  realization of a ground state quantum rotor}

\author{Igor Kuzmenko$^{1,2}$, Tetyana Kuzmenko$^1$,
  Y. Avishai$^{1,4}$ and Y. B. Band$^{1,2,3,4}$}

\affiliation{
  $^1$Department of Physics,
  $^2$Department of Chemistry,
  $^3$Department of Electro-Optics, and
  $^4$The Ilse Katz Center for Nano-Science,
  Ben-Gurion University of the Negev,
  Beer-Sheva 84105, Israel
  }

\date{\today}

\begin{abstract}
In a cold atom gas subject to a 2D spin-dependent optical lattice potential with hexagonal symmetry, trapped atoms execute circular motion around the potential minima.  Such atoms are elementary quantum rotors.  The theory of such quantum rotors is developed.  Wave functions, energies, and degeneracies are determined for both bosonic and fermionic atoms, and magnetic dipole transitions between quantum rotor states are elucidated. Quantum rotors in optical lattices with precisely one atom per unit cell can be used as extremely high sensitivity rotation sensors, accelerometers, and magnetometers.

\end{abstract}


\maketitle

\section{Introduction} \label{Sec:Intro}

A quantum mechanical system in which the motion of a particle is constrained 
to a circular ring (or a multi-dimensional spherical shell) is an
elementary  quantum rotor (QR) \cite{Sachdev}. References \cite{Sachdev, 
wiki_Quantum-rotor} state that ``elementary QRs do not exist in nature''. 
Here we consider both bosonic and fermionic cold atoms subject
to  a 2D spin-dependent optical lattice potential (SDOLP) of hexagonal
symmetry, and show that the trapped atoms behave as elementary QRs.
We demonstrate that QRs with {\em singly occupied sites} (so deleterious 
spin relaxation effects are thereby suppressed) can be used as a high 
accuracy rotation sensor, accelerometer, and magnetometer.

Quantum rotors have been formed using Laguerre-Gaussian (LG) beams 
\cite{Allen_92, Clifford_98}, e.g., see Refs.~\cite{Amico, Kumar}.
However, these references mostly considered the many-body
prosperities of cold atomic clouds in these beams and were not
focused on {\em elementary} quantum rotors.  In contrast, our interest 
is in singly occupied sites of SDOLPs so that interactions between QR atoms are
suppressed.  Hence the systems we consider are elementary QRs.  
Moreover, here we show that these QRs can be used as high accuracy rotation sensors, 
accelerometers, and magnetometers, and this not considered in Refs.~\cite{Amico, Kumar}.

The outline of the paper is a follows.  In Sec.~\ref{Sec:Model} we present the model of
the quantum rotors in the SDOLP.  Section \ref{SubSec:2D_isotropic} presents the
2D isotropic approximation for the SDOLP and Sec.~\ref{SubSec:Exotic} discusses the
exotic properties of these QRs.  The stimulated Raman spectroscopy used to probe the
QRs is considered in Sec.~\ref{Sec:SRS}, and in Sec.~\ref{SubSec:Raman} we focus on far-off-resonance Raman transitions between the ground state levels $n = 0, \zeta = 1/2$ and $n = 0, \zeta = -1/2$.  Section \ref{Sec:Magnetometer} explains how to use the QRs in the SDOLP with singly occupied sites as a high precision magnetometer, Sec.~\ref{Sec:Rotation Sensor} discusses the use of these QRs as rotation sensors and Sec.~\ref{Sec:Accelerometer} as accelerometers. 
In Sec.~\ref{Sec:Accuracy} we provide estimates of the accuracy of measurements
of rotation, acceleration and magnetic field using QRs, and in Sec.~\ref{Sec:shot-noise} 
we calculate the uncertainty due to shot noise in the Stokes and pump pulses. 
Finally, a summary and conclusion are presented in Sec.~\ref{Sec:Summary}.
In order to clarify some of the ideas presented in the main text, we provide 
a number of Appendices with background material and further details that 
substantiate the material presented in the main text.  Specifically, Sec.~\ref{Sec:OLP} 
provides additional details regarding SDOLP, and Sec.~\ref{Sec:Isotropic} presents
further details on the isotropic approximation for the potential near its minima.
The QR areal probability density is studied in Sec.~\ref{Sec:probability}.  
Section~\ref{Sec:Semiclassical} provides a semiclassical description of the quantum QR.  
Finally, Sec.~\ref{Sec:Distinguishing} discusses a method of distinguishing between the 
effects of rotation, acceleration and external magnetic field on the QR.

\section{The Model}  \label{Sec:Model}

\begin{figure}
\centering
  \includegraphics[width=60 mm,angle=0]
   {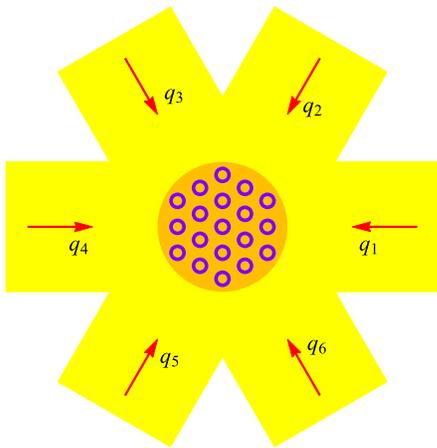}
 \caption{
   Laser beams (yellow) with wavevectors $\mbfq_n$ in the $x$-$y$ plane generate an
   optical lattice potential with hexagonal symmetry.
   The orange disk in the center shows the physical region in which the optical
   lattice is located.}
 \label{Fig-opt-latt-triangle}
\end{figure}

Consider alkaline atoms trapped in the $x$-$y$ plane by a SDOLP
\cite{Lin-Nature-2011,%
SOI-PRL-04,Galitski-Nature-2013,%
Juzeliunas-PRA-2010,Liu-PRL-2014,%
Zhang-PRL-2010,SOI-EuroPhysJ-13}.
As shown schematically in Fig.~\ref{Fig-opt-latt-triangle}, an optical
lattice potential with hexagonal symmetry \cite{other-symmetry}
is formed by 6 coherent laser beams of wavelength
$\lambda_0=\frac{2 \pi}{q_0}$ and wave-vectors
$\mbfq_n=-q_0(\cos(n\pi/3),\sin(n\pi/3))$ with $n=1,2,3,\ldots,6$.
The resultant electric field is $\mbfE(\mbfr,t)=(\mbfE({\bf r})
e^{-i \omega_0 t} + $ c.c.)/2 with amplitude $E_0$ and space dependent part
\begin{eqnarray}
  \mbfE(\mbfr) &=&
  E_0
  \sum_{n=1}^{6}
  \boldsymbol\xi_n
  e^{i \mbfq_n \mbfr},
  \label{Ej-triangle-def}
\end{eqnarray}
where the polarization vectors ${\boldsymbol \xi}_n$ are
\begin{eqnarray}
  &&
  \boldsymbol\xi_n =
  \bigg\{
       \sqrt{1-\beta^2}~
       \mbfe_z + \frac{\beta}{q_0}
       \mbfq_n \times \mbfe_z
  \bigg\}.
  \label{xi-vectors-triangle-def}
\end{eqnarray}
The electric field (\ref{Ej-triangle-def}) is a linear combination of standing waves with
in-plane and out-of-plane linear polarization with real mixing parameter $0<\beta<1$ (in 
what follows we take $\beta=1/\sqrt{2}$).  It generates an effective SDOLP experienced by the atoms.

The quantum states of the trapped atoms are described by the electronic angular
momentum $J$, the nuclear spin $I$, and the total internal atomic
angular momentum quantum number $F$ (${\bf F} = {\bf J} + {\bf I}$).  As we shall see, 
atoms with $F \ne 0$ rotate in a closed circular ring in the $x$-$y$ plane around
local minima of the scalar optical lattice potential.  The corresponding orbital angular momentum
operator is denoted by ${\boldsymbol \ell}$.  The projections of ${\bf F}$, ${\boldsymbol \ell}$ and the
total angular momentum of the QR, ${\boldsymbol {\mathcal{L}}} = {\bf F} + {\boldsymbol \ell}$,
on the $z$-axis are $f$, $m$ and $\zeta = f + m$ respectively.  For bosonic (fermionic) atoms 
$\zeta$ is an integer (half-integer).  Generically, the optical potential is not diagonal in 
$F$ nor in $f$ \cite{SOI-EuroPhysJ-13}, for further details see Appendix \ref{Sec:OLP}.
However, when the off-diagonal elements in $F$ are much smaller than the atomic
hyperfine splitting, the mixing of atomic energy levels with different
quantum numbers $F$ can be neglected.

For $J=1/2$, the optical lattice Stark interaction Hamiltonian is calculated 
as the second-order ac Stark shift.  In the hyperfine basis it takes the form
\cite{SOI-EuroPhysJ-13, hyperfine} (see also Appendix \ref{apend-OP-J=1/2} for details),
\begin{equation}    \label{HStark=V+BF}
  H_{\mathrm{Stark}}(\mbfr) = V(\mbfr) 1 - \mbfB(\mbfr) \cdot \mbfF,
\end{equation}
where ${\bf 1}$ is the $(2F+1)$$\times$$(2F+1)$ unity matrix.  The scalar optical 
potential $V(\mbfr)$ and fictitious magnetic field $\mbfB(\mbfr)$ \cite{CCT} (which 
has units of energy) are
\begin{subequations}
\begin{eqnarray} 
  V(\mbfr) &=&
  -\frac{\alpha_0(\omega_0)}{4}~ \mbfE^{*}(\mbfr)
      \cdot
      \mbfE(\mbfr),
        \label{V-def} \\
  \mbfB(\mbfr) &=&
  \frac{i \alpha_1(\omega_0)}{4 (2 I+1)}~ [
      \mbfE^{*}(\mbfr)
      \times
      \mbfE(\mbfr) ],
      \label{B-def}
\end{eqnarray}
  \label{subeqs-V-B-def}
\end{subequations}
where $\alpha_0(\omega_0)$ and $\alpha_1(\omega_0)$
are scalar and vector polarizabilities of the atom
\cite{SOI-EuroPhysJ-13,Li-polarizability-2017}. 
For $J > 1/2$, a tensor term is also present in Eq.~(\ref{W=V+BF})
(see Ref.~\cite{SOI-EuroPhysJ-13}).

Both $V(\mbfr)$ and $\mbfB(\mbfr)$ are periodic,
$V(\mbfr) = V(\mbfr+ m_1 \mbfa_1 + m_2 \mbfa_2)$
and
$\mbfB(\mbfr) = \mbfB(\mbfr+ m_1 \mbfa_1 + m_2 \mbfa_2)$,
where $m_1$ and $m_2$ are integers, the lattice vectors are
$\mbfa_j=\lambda_0 (\mbfe_x \sin\theta_j + \mbfe_y \cos\theta_j)$, $j=1,2$, and
$\theta_j=(-1)^{j}\pi/3$.
$V(\mbfr)$ has minima at $\mbfr_{\mathrm{min}}=m_1\mbfa_1+m_2\mbfa_2$.
$\mbfB(\mbfr)$ is a pseudovector \cite{SOI-EuroPhysJ-13}, and 
changes sign under time reversal (but the QR Hamiltonian is
time reversal invariant).  $V(\mbfr)$ and $\mbfB(\mbfr)$ are plotted versus 
${\bf r}$ in Fig.~\ref{Fig-V-B-tr} in Appendix \ref{apend-OP-J=1/2}.

The loss of atoms from the near-detuned SDOLP due to excited
state spontaneous emission can be phenomenologically taken into account by including
an imaginary contribution to the energy denominator of the second order ac Stark shift,
see Eq.~(\ref{polar-reduced-def}) of Appendix \ref{Sec:OLP}.  The loss rate is given by 
\begin{eqnarray}
  \Gamma(\mbfr)  &=& \frac{V(\mbfr) \gamma_{\e}}{2 \hbar \Delta} .
  \label{Gamma-detune}
\end{eqnarray}
where $\Delta$ is the detuning of the optical frequency from resonance 
and $\gamma_{\e}$ is the inverse lifetime of the excited electronic state. 
As we shall soon see, the loss rate $\Gamma(\mbfr)$ will affect the accuracy 
of the sensors considered below.

Candidates for observing QR states include the fermions $^2$H, $^6$Li, $^{40}$K, and 
the bosons $^7$Li, $^{23}$Na, $^{39}$K, $^{85}$Rb and $^{87}$Rb.
All of these have nonvanishing $F$ in their
ground electronic state \cite{hyperfine-H-He-PhysLett-2002}.
Recoil temperatures $T_{0}={\mathcal{E}}_{0}/k_B$
for some of these fermionic and bosonic species are listed in Table~\ref{tab-recoil}.

\begin{table} 
\caption{Recoil temperatures $T_{0}={\mathcal{E}}_{0}/k_B =
\hbar^2 q_{0}^{2}/(2 M k_B)$ for some atoms.}
\label{tab-recoil}
\centering.
\begin{tabular}{|c||c|}
  \hline
  fermions & bosons
  \\
  \hline
  \begin{tabular}{c|c}
    atom & $T_0$ ($\mu$K)
    \\
    \hline
    ${}^{2}{\text{H}}$ &
    321.7
    \\
    \hline
    ${}^{6}{\text{Li}}$ &
    3.536
    \\
    \hline
    ${}^{40}{\text{K}}$ &
    0.404
  \end{tabular}
  &
  \begin{tabular}{c|c}
    atom & $T_0$ ($\mu$K)
    \\
    \hline
    ${}^{7}{\text{Li}}$ &
    3.031
    \\
    \hline
    ${}^{23}{\text{Na}}$ &
    1.197
    \\
    \hline
    ${}^{39}{\text{K}}$ &
    0.414
  \end{tabular}
  \\
  \hline
\end{tabular}
\end{table}

\subsection{2D Isotropic approximation}  \label{SubSec:2D_isotropic}
We assume hereafter that the depth of $V(\mbfr)$ at
the minimum positions exceeds the recoil energy
${\mathcal{E}}_{0}=\hbar^2 q_{0}^{2}/(2 M)$, where $M$ is the atomic mass, and the low-energy
atomic states are trapped and localized near these minima. For hexagonal symmetry, 
the scalar potential near the minimum at $\mbfr=(0,0)$ is well-approximated by 
$V({\mathbf{r}}) \approx \tilde{V}(r)$, where
\begin{eqnarray}
  {\tilde V}(r) = -\frac{V_0}{6}~
  \Bigg[2 + 3
       J_0\bigg(\frac{2 \pi r}{\lambda_0}\bigg)+
       J_0\bigg(\frac{2 \sqrt{3} \pi r}{\lambda_0}\bigg)
  \Bigg],
  \label{V-isotropic}
\end{eqnarray}
and $V_0 = \frac{9}{2} \alpha_0(\omega_0) E_{0}^{2}$. The fictitious magnetic field near this minimum 
can be approximated by $\mbfB(r) \approx \tilde{B}(r) \mbfe_r$ where
\begin{eqnarray}
  \tilde{B}(r) &=&
  \frac{B_0}{3 (2I+1)}~
  \Bigg[
       J_1
       \bigg(
            \frac{2 \pi r}{\lambda_0}
       \bigg)+
       J_1
       \bigg(
           \frac{4 \pi r}{\lambda_0}
       \bigg)+
  \nonumber \\ &&
       \sqrt{3}~
       J_1
       \bigg(
            \frac{2 \sqrt{3} \pi r}{\lambda_0}
       \bigg)
  \Bigg].
  \label{B-isotropic}
\end{eqnarray}
Here $B_0 = \frac{9}{2} \alpha_1(\omega_0) E_{0}^{2}$, $J_n(\rho)$ is the Bessel function 
of order $n$, and the unit vector $\mbfe_r \equiv \hat{\bf{r}}$.  The dependence of 
$V_0$ and $B_0$ on the detuning of $\omega_0$ from resonance is discussed in
Appendix \ref{apend-OP-J=1/2}.

The (fictitious) Zeeman interaction ${\bf B}({\bf r}) \cdot {\bf F}$ is proportional to $F_r = {\bf F} \cdot \mbfe_r$
[since ${\bf B}({\bf r})$ is radial] and does not commute with the operators $F_z$ or $\ell_z$ ($=-i\partial_{\phi}$),
but commutes with ${\mathcal{L}}_z=F_z+\ell_z$.
Hence, $f$ and $m$ are not individually
good quantum numbers of the total Hamiltonian but $\zeta$,
the eigenvalue of the operator ${\mathcal{L}}_z$, is.
The wave functions $\Psi_{n,\zeta}(\mbfr)$ and energy
levels $\epsilon_{n,\zeta}$ of the trapped atoms satisfy
the Schr\"odinger equation \cite{band_structure},
\begin{eqnarray}
  \bigg[
       -\frac{\hbar^2 \nabla^2}{2 M}+
       {\tilde V}(r)-
       \tilde{B}(r)F_r
       -\epsilon_{n,\zeta}
  \bigg] \Psi_{n,\zeta}(\mbfr)
  = 0 ,
  \label{eq-Schrodinger-def}
\end{eqnarray}
and the functions $\Psi_{n,\zeta}(\mbfr)$ can be expanded as follows:
\begin{eqnarray}
  \Psi_{n,\zeta}(\mbfr) &=&
  \frac{1}{\sqrt{2 \pi r}}
  \sum_{\sigma}
  \psi_{n,\zeta,\sigma}(r)
  e^{i \zeta \phi}
  \chi_{\sigma}(\phi).
  \label{WF-vs-WF-f-m}
\end{eqnarray}
Here $\chi_{\sigma}(\phi)$ are spinor eigenfunctions of $F_r$, i.e.,
$F_r \chi_{\sigma}(\phi) = \sigma \chi_{\sigma}(\phi)$,
($\sigma$ should not be confused with $f$, the eigenvalue of $F_z$).

\begin{figure}
\centering
  {\includegraphics[width=0.7\linewidth,angle=0]
   {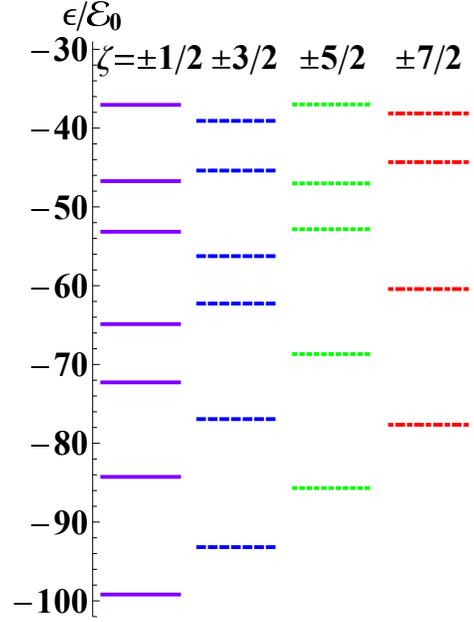}}
 \caption{
   Fermionic QR energy levels for $F = 1/2$ with
   $V_0=100 \, {\mathcal{E}}_{0}$ and
   $B_0=180 \, {\mathcal{E}}_{0}$.
   The solid purple, dashed blue,
   dotted green and dot-dashed red
   lines correspond to energy
   levels with $\zeta = \pm1/2$, $\pm3/2$,
   $\pm5/2$, $\pm7/2$.
   The energy levels with
   $n=0,1,\ldots$, go from bottom to top.}
 \label{Fig-Sp}
\end{figure}

Explicitly, for $F=1/2$ (e.g., ${}^6$Li), $\sigma= \pm 1/2$, and $\chi_{\sigma}(\phi)$ is given by
$$
  \chi_{\sigma = \pm 1/2}(\phi)
  =
  \frac{1}{\sqrt{2}}~
  \Big\{
      e^{-\frac{i \phi}{2}}
      \chi_{\ua}^{(z)} \pm
      e^{\frac{i \phi}{2}}
      \chi_{\da}^{(z)}
  \Big\},
$$
where $\chi_{f}^{(z)}$ are eigenfunctions
of $F_z$ with eigenvalues $f$.
Inserting the expansion (\ref{WF-vs-WF-f-m}) into
Eq.~(\ref{eq-Schrodinger-def}) we obtain,
\begin{subequations}    \label{eq-Schrodinger-radial}
\begin{eqnarray}
    &&\bigg[
       -\frac{\hbar^2}{2 M}
       \frac{\partial^2 }{\partial r^2} +
       {\tilde V}(r) -
       \frac{1}{2}~{\tilde{B}}(r) + \zeta^2 C(r)
       - \epsilon_{n,\zeta}
  \bigg] \times \nonumber \\
  &&   \psi_{n,\zeta,1/2}(r)   =
  \zeta C(r)
  \psi_{n,\zeta,-1/2}(r),
  \label{eq-Schrodinger-up}
  \\
  && \bigg[
       -\frac{\hbar^2}{2 M}
       \frac{\partial^2 }{\partial r^2} +
       {\tilde V}(r) +
       \frac{1}{2}~{\tilde{B}}(r) +
       \zeta^2 C(r) -  \epsilon_{n,\zeta}
  \bigg] \times \nonumber \\
   && \psi_{n,\zeta,-1/2}(r)  =
  \zeta C(r)
  \psi_{n,\zeta,1/2}(r),
  \label{eq-Schrodinger-dn}
\end{eqnarray}
\end{subequations}
where the off-diagonal spin-flipped terms are proportional to
$C(r) = \hbar^2/(2 M r^2)$.  The numerical calculations
presented below will be for the fermionic case with $F=1/2$, assuming 
that $V_0,|B_0| \gg {\mathcal{E}}_{0}$.
The inequality $V_0 \gg {\mathcal{E}}_{0}$ means that the potential
wells are deep and the tunneling probability of the atoms between the wells is small.  
The wave functions and the eigen-energies $\epsilon_{n,\zeta}$ of the trapped atoms 
are computed by solving the Sturm-Liouville system of equations (\ref{eq-Schrodinger-radial}) with $V_0 = 100 {\mathcal{E}}_{0}$, $B_0 = 180 {\mathcal{E}}_{0}$.
The resulting energies $\epsilon_{n,\zeta}$ are shown in Fig.~\ref{Fig-Sp}.
For the fermionic case, all the energy levels are two-fold degenerate,
$\epsilon_{n,\zeta}=\epsilon_{n,-\zeta}$.  The off-diagonal terms
$\zeta C(r) \psi_{n,\zeta,\overline\sigma}(r)$ (where $\overline\sigma=-\sigma$) 
in (\ref{eq-Schrodinger-radial}) are a weak perturbation to the ground-state
energy level, but not so for the excited states.
The degenerate fermionic ground state has quantum numbers $(n,\zeta)=(0,\pm 1/2)$, 
and energy $\epsilon_{0,\pm 1/2} = -99.196 \, {\mathcal{E}}_{0}$.
The lowest energy excited states are orbital excitations with quantum
numbers $(n,\zeta)=(0,\pm 3/2)$ and the radial excitations with quantum
numbers $(1,\pm 1/2)$, see Fig.~\ref{Fig-Sp}.  The corresponding excitation energies 
are,
\begin{subequations}   \label{subeqs-energies-orbit-spin-harmonic}
\begin{eqnarray}
  \epsilon_{0,\frac{3}{2}}-
  \epsilon_{0,\frac{1}{2}}
  &=&
  6.011~
  {\mathcal{E}}_{0},
  \label{energy-orbit}
  \\
  \epsilon_{1,\frac{1}{2}}-
  \epsilon_{0,\frac{1}{2}}
  &=&
  14.93~
  {\mathcal{E}}_{0},
  \label{energy-harmonic}
 \end{eqnarray}
\end{subequations}
For temperatures $T \ll T_{\mathrm{orbit}} \equiv
(\epsilon_{0,\frac{3}{2}}-\epsilon_{0,\frac{1}{2}})/k_B$,
the trapped atoms are in their ground state.
$T_{\mathrm{orbit}}$ is of order $T_{0}$, see Eq.~(\ref{energy-orbit}).

\begin{figure}
\centering
  \includegraphics[width=0.9 \linewidth]
   {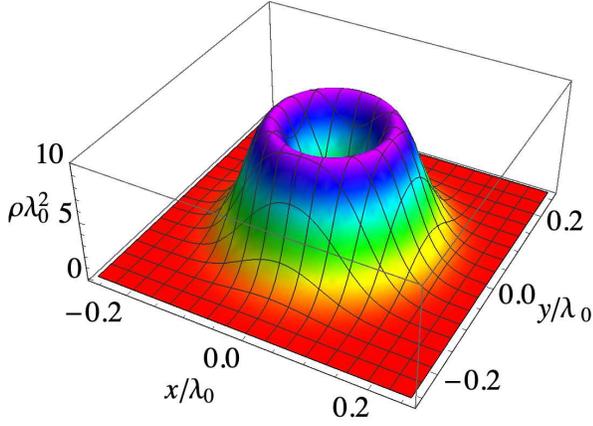}
 \caption{
   The ground state areal probability density (\ref{density-low-T})
   versus $\mbfr=(x,y)$ for a $F = 1/2$ fermionic atom.
   }
 \label{Fig-density}
\end{figure}

The ground state areal probability density,
\begin{eqnarray}
  \rho(r) &=& \big| \Psi_{0,1/2}({\bf r}) \big|^{2},
  \label{density-low-T}
\end{eqnarray}
depends only on $r$ and not on $\phi$ since the optical lattice potential is nearly
isotropic about the potential minima.  $\rho(r)$ has a maximum at
$r=r_0=0.068 \, \lambda_0$, and decays with $|r-r_0|$, as shown in Fig.~\ref{Fig-density}.
Thus, the atom is confined near a circular ring of radius $r_0$ (hence it is a QR \cite{Sachdev}). 
$\rho(r)$ vanishes linearly with $r$ as $r \to 0$.
The QR ground state density can be observed as described in Ref.~\cite{Bloch-position-Nature-2011}.

\subsection{Exotic properties of the QR state}  \label{SubSec:Exotic}

The following exotic expectation values are obtained for the QR:
\begin{eqnarray}
  \big\langle n,\zeta \big| \ell_z \big| n,\zeta \big\rangle &=&  \zeta -
  \big\langle n,\zeta \big| F_z \big| n,\zeta \big\rangle ,
  \label{ME-Lz}
  \\
  \big\langle n,\zeta \big| \ell_{z}^{2} \big| n,\zeta \big\rangle &=&
  \zeta^2 - \big\langle n,\zeta \big| 2 \zeta F_z - F_{z}^{2} \big| n,\zeta \big\rangle .
  \label{ME-Lz-2}
\end{eqnarray}
Recall  that $\Psi_{n,\zeta}(\mbfr)=\la {\bf r}\vert n,\zeta \ra$ are eigenfunctions
of ${\mathcal{L}}_{z}=\ell_z+F_z$ with eigenvalue $\zeta$, however $\Psi_{n,\zeta}(\mbfr)$ 
is not an eigenfunction of $F_z$ or $F_{z}^{2}$, and
$\la n,\zeta \vert F_z^2 \vert n,\zeta \ra \ne 0$.

For fermionic or bosonic QRs, the wave functions $\Psi_{n,\zeta}(\mbfr)$ in
Eq.~(\ref{WF-vs-WF-f-m}) are expressed as  sums of products of
spatial wave functions $\psi_{n,\zeta,\sigma}(r)e^{i \zeta \phi}$
and spin wave functions $\chi_{\sigma}(\phi)$.
They have unusual symmetry
relations under rotation through an angle $2\pi$ around the $z$ axis.
The angular part of the spatial wave function, satisfies
$e^{i \zeta (\phi+2 \pi)} = \pm e^{i \zeta \phi}$
(upper sign for bosons and lower sign  for fermions),
and the spin wave function satisfies
$e^{2 \pi i {\mathcal{L}}_z} \chi_{\sigma}(\phi) = \chi_{\sigma}(\phi)$.

For bosonic QRs, $\zeta$ and $\sigma$ are integers, so the angular part
of the spatial wave functions, $e^{i \zeta \phi}$, and the spin
wave functions satisfy the properties,
$e^{i \zeta (\phi+2 \pi)} = e^{i \zeta \phi}$ and
$e^{2 \pi i {\mathcal{L}}_z} \chi_{\sigma}(\phi) = \chi_{\sigma}(\phi)$.
The non-degenerate ground state $\vert n =0, \zeta = 0 \ra $ is such that 
$\la 0,0\vert F_z \vert 0,0 \ra=0$, but $\la 0,0\vert F^2_z \vert 0,0 \ra \ne 0$, 
and therefore $\la 0,0\vert \ell^2_z \vert 0,0 \ra \ne 0$.
The spin-excited QR states have $\zeta \ne 0$ and are doubly degenerate
(see Eq.~(1.23) in Ref.~\cite{Sachdev}).  All states have an areal
density which vanishes at $r = 0$.

For fermionic QRs, $\zeta$ and $\sigma$ are half-integer, and all 
states (including the ground state) are doubly degenerate.  
The ground state 
has $n=0$ and $\zeta=\pm 1/2$, and is an {\em exotic QR} since 
it is {\em two-fold degenerate} with finite orbital angular momentum. 
The  expectation values of $\ell_z$ and $\ell_{z}^{2}$ are non-zero. For $F=1/2$, 
$
  \big\langle
      n,\zeta
  \big|
      \ell_z
  \big|
      n,\zeta
  \big\rangle
  =
  \frac{\zeta}{2} -
  \beta^{z}_{n,\zeta}$,
and  $\big\langle
      n,\zeta
  \big|
      \ell_{z}^{2}
  \big|
      n,\zeta
  \big\rangle
  =  \frac{1}{2} -
  \big|\beta^z_{n,\zeta}\big|$,
where
\begin{equation}  \label{eq:beta}
\beta_{n,\zeta}^{z} =
  \frac{1}{2}
  \sum_{\sigma}
  \int
  \psi_{n,\zeta,\sigma}
  \psi_{n,\zeta,-\sigma} dr.
\end{equation}

A striking consequence of the above analysis, which will be substantiated below,
is that ground state QRs with 
precisely one atom per site can serve as a rotation sensor,
an accelerometer, and a magnetometer.  For these applications and for the study
of QRs in general, radio wave spectroscopy and Raman spectroscopy are valuable tools.

\section{QR Stimulated Raman Spectroscopy}  \label{Sec:SRS}

Consider Raman transitions between the QR states $|n = 0, \zeta = 1/2\rangle$ and
$|n = 0, \zeta = -1/2\rangle$ that are split by an energy $\hbar \Delta_{QR}$
due to the presence of an external magnetic field, and/or rotation, and/or 
in-plane acceleration.
We explicitly consider far-off-resonance radio wave transitions that are red-detuned
from the $F=3/2$ atomic hyperfine state by a large detuning $\Delta_{\mathrm{hf}}$,
$\Delta_{\mathrm{hf}} < 0$ and $|\Delta_{\mathrm{hf}}| \gg \gamma_{\mathrm{hf}}$,
where $\gamma_{\mathrm{hf}}$ is the decay rate of the $F=3/2$ hyperfine
state (see Fig.~\ref{Fig-Raman-Rabi} in Sec.~\ref{SubSec:Raman}).  
Far-off-resonance stimulated Raman scattering can be treated as a two-level 
system with a generalized Raman Rabi frequency, $\Omega_g = 
\frac{\Omega_p \Omega_s}{\Delta_{\mathrm{hf}}}$.
The pump laser has frequency $\omega_p$ and Rabi frequency $\Omega_p$, and
takes the system up from the lower of the two QR states to a virtual intermediate
state, and the Stokes radiation has frequency $\omega_s$ and Rabi frequency $\Omega_s$,
and takes the system down from the virtual intermediate state to the upper of the 
two QR states.
Let $\delta = \omega_p - \omega_s - \Delta_{QR}$ be the detuning from Raman resonance.
The dressed-state \cite{Cohen-Tannoudji} complex Hamiltonian in the
2-level $|n = 0, \zeta = \pm 1/2\rangle$ manifold, which incorporates decay of the QR states, 
can be written as \cite{Sokolov_92}
\begin{equation} \label{eq:H_raman}
H_{\mathrm{Raman}} = \hbar \left(
    \begin{array}{cc}
      0 - i \Gamma_{0,1/2}/2 &
      \Omega_g/2
      \\
      \Omega_g/2 &
      \delta - i \Gamma_{0,-1/2}/2
    \end{array}
  \right) .
\end{equation}
Symmetry requires that the loss rates $\Gamma_{0,\pm 1/2}$ of the two QR states 
due to loss of atoms from the SDOLP be equal, $\Gamma_{0,1/2} = \Gamma_{0,-1/2}$; they are
given by (see Sec.~\ref{SubSec:Raman} below for details):
\begin{eqnarray}
  \Gamma_{0,1/2} &=&
  \frac{\gamma_{\e}}{2 \hbar \Delta}
  \sum_{\sigma}
  \bigg\{
       \int
       \tilde{V}(r)
       \psi_{0,1/2,\sigma}^{2}(r)
       dr
  \nonumber \\ && +
       \frac{{\mathrm{sign}}(\sigma)}{2}
       \int
       \tilde{B}(r)
       \psi_{0,1/2,\sigma}^{2}(r)
       dr
  \bigg\}.
  \label{Gamma-rate}
\end{eqnarray}
Here $\Delta$ is the detuning of the laser beams generating
the SDOLP and $\gamma_{\e}$ is the spontaneous emission decay rate of the 
${}^{2}$P$_{3/2}$ excited state.  The difference of the eigenvalues of 
$H_{\mathrm{Raman}}$, $\tilde\Omega=\sqrt{\delta^2 + \Omega_{g}^{2}}$, is 
independent of $\Gamma_{0,1/2}$.

\subsection{Far-Off-Resonance Raman Transitions Between $n = 0, \zeta = 1/2$ and $n = 0, \zeta = -1/2$}
  \label{SubSec:Raman}

Far-Off-Resonance Raman transitions between the QR states $|n = 0, \zeta = 1/2\rangle$
and $|n = 0, \zeta = -1/2\rangle$ that are split by an energy $\hbar \Delta_{QR}$
is described by the model Hamiltonian (\ref{eq:H_raman}).
Here we derive Hamiltonian (\ref{eq:H_raman}), its eigenvalues and eigenfunctions.

\begin{figure} [htb]
\centering
  \includegraphics[width=0.8\linewidth,angle=0]
   {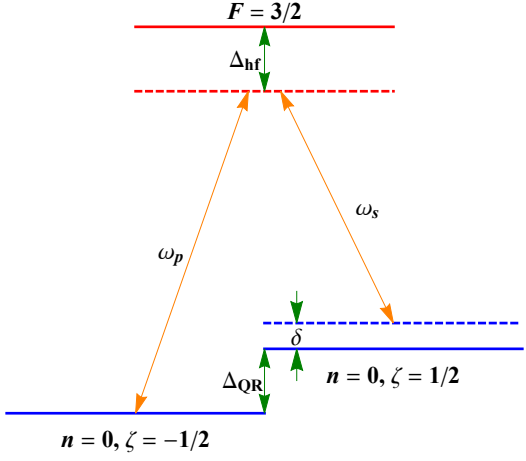}
   \caption{Schematic representation of the energy levels and photon energies for the 
   far-off-resonance Raman transition between the split QR ground states.}
 \label{Fig-Raman-Rabi}
\end{figure}

Consider Raman transitions between the QR states $|n = 0, \zeta = 1/2\rangle$ and 
$|n = 0, \zeta = -1/2\rangle$ that are split by an energy $\hbar \Delta_{QR}$ by the presence of a rotation, or an external magnetic field, or an in-plane acceleration.
When the red detuning $\Delta_{\mathrm{hf}}$ from the $F=3/2$ hyperfine state is
large enough, $|\Delta_{\mathrm{hf}}| \gg \gamma_{\mathrm{hf}}$ and 
$\Delta_{\mathrm{hf}} < 0$, where $\gamma_{\mathrm{hf}}$ is the decay rate of the 
$F = 3/2$ hyperfine state, see Fig.~\ref{Fig-Raman-Rabi}, the off-resonance intermediate 
$F = 3/2$ state can be eliminated and the Raman process can be treated as a 2-level problem.
The resultant generalized Raman Rabi frequency is given by $\Omega_g = 
\frac{\Omega_p \Omega_s}{\Delta_{\mathrm{hf}}}$.  If we denote the energy 
difference between the two levels as $\hbar \Delta_{QR}$,
the detuning from Raman resonance is $\delta = \omega_p - \omega_s - \Delta_{QR}$.
The dressed-state 2-level non-Hermitian Hamiltonian \cite{Cohen-Tannoudji} which 
incorporates decay of the QR states can be written as \cite{Sokolov_92}
\begin{equation} \label{eq:H_raman-gen}
  H_{\mathrm{Raman}} = {\mathcal{H}}_{\mathrm{Raman}}
   - \frac{i \hbar}{2} ~ {\mathcal{G}},
\end{equation}
where ${\mathcal{H}}_{\mathrm{Raman}}$ and ${\mathcal{G}}$ are 2$\times$2 Hermitian
matrices; the Hermitian part ${\mathcal{H}}_{\mathrm{Raman}}$ in (\ref{eq:H_raman-gen}) is
\begin{eqnarray}  \label{eq-H-Raman-Hermitian}
  {\mathcal{H}}_{\mathrm{Raman}} &=&
  \hbar \left(
    \begin{array}{cc}
      0 &
      \Omega_g/2
      \\
      \Omega_g/2 &
      \delta
    \end{array}
  \right) .
\end{eqnarray}
It acts in the two dimensional Hilbert states spanned by $(1,0)^{T}=|0,1/2\rangle$ 
and $(0,1)^{T}=|0,-1/2\rangle$.  
The anti-Hermitian part originates from the decay rate in the optical lattice,
\begin{eqnarray}
  \Gamma(\mbfr) &=&
  \frac{\gamma_{\e}}{2 \hbar \Delta}~
  \Big\{
      \tilde{V}(r) +
      F_r
      \tilde{B}(r)
  \Big\},
  \label{decay-rate-opt-latt}
\end{eqnarray}
where $\gamma_{\e}$ is the decay rate of the excited ${}^{2}$P$_{3/2}$ state,
and $\Delta$ is the detuning of the laser frequency from the resonant frequency.
When $|\Delta| \gg \gamma_{\e}$, we can apply perturbation theory and
write the anti-Hermitian part the Hamiltonian (\ref{eq:H_raman-gen}) as
\begin{eqnarray*}
  {\mathcal{G}}_{\zeta,\zeta'} &=&
  \int
  \Psi_{0,\zeta}^{\dag}(\mbfr)
  \Gamma(\mbfr)
  \Psi_{0,\zeta'}(\mbfr)~
  d^2\mbfr,
\end{eqnarray*}
where $\Psi_{n,\zeta}(\mbfr)$ are the wave functions of the QR
given by Eq.~(10).  Note that the atom in the quantum state with $\zeta=1/2$ 
or $-1/2$ orbits clockwise or counterclockwise around the minimum
of the lattice potential. The decay rate (\ref{decay-rate-opt-latt})
is isotropic. Therefore taking into account the symmetry of
the wave function
$$
  \Psi_{n,-\zeta}(r,\phi) =
  F_x
  \Psi_{n,\zeta}(r,-\phi),
$$
(which is true for $F=1/2$ atoms), we can write
\begin{eqnarray}
  {\mathcal{G}}_{\zeta,\zeta'} =
  \Gamma_{0,1/2}~
  \delta_{\zeta,\zeta'},
  \label{eq-H-Raman-anti-Hermitian}
\end{eqnarray}
where $\Gamma_{0,1/2}$ is given by Eq.~(\ref{Gamma-rate}).
Eqs.~(\ref{eq:H_raman-gen}), (\ref{eq-H-Raman-Hermitian}) and
(\ref{eq-H-Raman-anti-Hermitian}) yield Eq.~(\ref{eq:H_raman}).

Another source of uncertainty is spontaneous decay of
the $F = 3/2$ hyperfine state, which gives a decay rate
\begin{eqnarray}
  \Gamma_{\mathrm{hf}} &=&
  \frac{\gamma_{\mathrm{hf}}}{\hbar \Delta_{\mathrm{hf}}}~
  \sum_{\sigma}
  \int
  \bigg\{
       \tilde{V}(r) +
       \frac{{\mathrm{sign}}(\sigma)}{2}
       \tilde{B}(r)
  \bigg\}
  \times \nonumber \\ && \times
  \psi_{0,1/2,\sigma}^{2}(r) dr,
  \label{Gamma-hf}
\end{eqnarray}
For $^6$Li atoms
in the ground state, $\gamma_{\mathrm{hf}} = 1.586 \times 10^{-17}~{\text{s}}^{-1}$
\cite{Landau-Lifshitz-4}. Comparing Eqs.~(\ref{Gamma-rate}) and (\ref{Gamma-hf}),
one concludes that
$$
  \frac{\gamma_{\mathrm{hf}}}{\Delta_{\mathrm{hf}}}
  ~\ll~
  \frac{\gamma_{\e}}{\Delta_{\e}},
$$
hence $\Gamma_{\mathrm{hf}} \ll \Gamma_{0,1/2}$. Thus,
in the following discussions we neglect $\Gamma_{\mathrm{hf}}$
since it is very small in comparison with $\Gamma_{0,1/2}$.

Eigenfunctions of the non-Hermitian Hamiltonian
(\ref{eq:H_raman}) are
\begin{subequations}
\begin{eqnarray}
  \big|
      \psi_{+}
  \big\rangle
  &=&
  {\mathcal{U}}
  \big|0,1/2\big\rangle +
  {\mathcal{V}}
  \big|0,-1/2\big\rangle,
  \label{WF-Raman-P}
  \\
  \big|
      \psi_{-}
  \big\rangle
  &=&
  {\mathcal{V}}
  \big|0,1/2\big\rangle -
  {\mathcal{U}}
  \big|0,-1/2\big\rangle,
  \label{WF-Raman-M}
\end{eqnarray}
  \label{subeqs-WF-Raman-PM}
\end{subequations}
where
\begin{eqnarray*}
  {\mathcal{U}} =
  \sqrt{\frac{\tilde\Omega + \delta}{2 \tilde\Omega}},
  \ \ \
  {\mathcal{V}} =
  \sqrt{\frac{\tilde\Omega - \delta}{2 \tilde\Omega}},
  \ \ \
  \tilde\Omega = \sqrt{\delta^2 + \Omega_{g}^{2}}.
\end{eqnarray*}
The corresponding eigenvalues are
$$
  \epsilon_{\pm} ~=~
  \frac{\hbar \big(\delta - i \Gamma_{0,1/2}\big)}{2}
  \pm
  \frac{\hbar \tilde\Omega}{2}.
$$
The difference $\epsilon_{+} - \epsilon_{-} = \hbar \tilde\Omega$
does not depend on $\Gamma_{0,1/2}$.

The time evolution of wave function of the QR with time, starting with the 
initial wave function $|0,1/2\rangle$ is specified by the time-dependent wave function
\begin{eqnarray}
  |\psi(t)\rangle &=&
  e^{- (\Gamma_{0,1/2} + i \delta)t/2}
  \times \nonumber \\ &&
  \bigg\{
       \bigg[
            \cos\bigg(\frac{\tilde\Omega t}{2}\bigg) +
            \frac{i \delta}{\tilde\Omega}
            \sin\bigg(\frac{\tilde\Omega t}{2}\bigg)
       \bigg]
       \big|0,1/2\big\rangle -
  \nonumber \\ &&
       \frac{i \Omega_g}{\tilde\Omega}
       \sin\bigg(\frac{\tilde\Omega t}{2}\bigg)~
       \big|0,-1/2\big\rangle
  \bigg\}.
  \label{WF-Raman-time-depend}
\end{eqnarray}
The probabilities $P_{0, \pm 1/2}(t)$ to find the QR in
the states $|0,\pm1/2\rangle$ are given by $P_{0, \pm 1/2}(t) =
\Big| \big\langle 0, \pm 1/2 \big| \psi(t) \big\rangle \Big|^{2}$.
Using Eq.~(\ref{WF-Raman-time-depend}), we find
\begin{subequations}    \label{subeqs-Probab-PM}
\begin{eqnarray}
  P_{0,1/2}(t) &=&
  \bigg[
       1-P_0\sin^2\bigg(\frac{\tilde\Omega t}{2}\bigg)
  \bigg]
  e^{-\Gamma_{0,1/2} t},
  \label{Probab-P}
\end{eqnarray}
\begin{eqnarray}
  P_{0,-1/2}(t) &=&
  P_0\sin^2\bigg(\frac{\tilde\Omega t}{2}\bigg)
  e^{-\Gamma_{0,1/2} t},
  \label{Probab-M}
\end{eqnarray}
\end{subequations}
where
$$
  P_0=
  \frac{\Omega_{g}^{2}}{\tilde\Omega^2} =
  \frac{\Omega_{g}^{2}}{\delta^2 + \Omega_{g}^{2}}.
$$
The Ramsey time-separated oscillating field method \cite{Ramsey_50} with Raman pulses 
\cite{Zanon} using radio-frequency pump and Stokes radiation
can be used to determine $\Delta_{QR}$, as we shall now show.

\begin{figure} [htb]
\centering
  \includegraphics[width=0.75 \linewidth,angle=0]
   {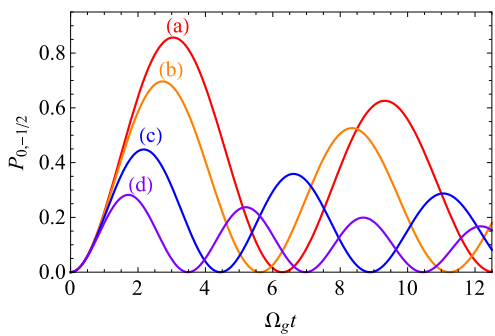}
 \caption{The probability $P_{0, -1/2}(t)$ in Eq.~(\ref{Probab-M}) to find the QR in 
   the state
   $|0,-1/2\rangle$ as a function of time for $\Gamma_{0,1/2} = 0.05 \, \Omega_g$,
   $\Omega_g = 1$ and different values of $|\delta|$:
   (a) $\delta=0$, (b) $|\delta| = 0.5 \, \Omega_g$, (c) $|\delta| = \Omega_g$ and 
   (d) $|\delta| = 1.5 \, \Omega_g$.}
 \label{Fig-Prob-Rabi}
\end{figure}

The probability $P_{0, - 1/2}(t)$ is plotted in Fig.~\ref{Fig-Prob-Rabi} for 
different values of $|\delta|$.  The amplitude $P_0$ of $P_{0,-1/2}(t)$ is maximal
when $\delta=0$. This is because $P_0$ is maximum when $\delta=0$ 
($\Delta_{QR} = \omega_p-\omega_s$), and becomes very small 
for weak stimulated Raman scattering, i.e., when $|\delta| \gg \Omega_g$.
Experimentally scanning $\omega_s$ and finding $\omega_p - \omega_s$ where $P_{0,-1/2}(t)$
is maximal yields $\Delta_{QR}$.

\subsection{Ramsey Separated Oscillating Field Method}  \label{SubSec:Ramsey}

\begin{figure} [htb]
\centering
\includegraphics[width=0.75 \linewidth]
{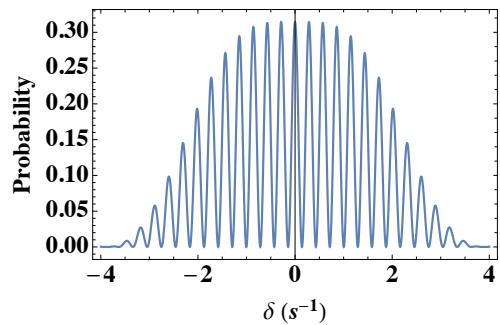}
\caption{Ramsey fringes in the probability $P_{|0,-1/2\rangle}$ at the
final time $T+2\tau_p$ plotted versus detuning $\delta$.  The splitting
$\Delta_{QR}$ is found by identifying where the detuning $\delta = \omega_p - 
\omega_s - \Delta_{QR} = 0$.}
\label{Fig_Ramsey_Fringe_QR}
\end{figure}

A preferable method of experimentally determining $\Delta_{QR}$ is to employ the Ramsey 
time-separated oscillating field method \cite{Ramsey_50} with Raman pulses \cite{Zanon}.
The QR initially in the ground state $|0,1/2\rangle$ is subjected to two sets of Raman pulses 
of duration $\tau_p$ separated by a delay time $T$, so the generalized Rabi frequency
becomes time-dependent:
\begin{equation}
\Omega_g(t) = \left\{ \begin{array}{ll} 
     \Omega_g & \mbox{if $0 \leq t \leq \tau_p$}\,, \\
     0 & \mbox{if $\tau_p < t < T+\tau_p$}\,, \\
     \Omega_g & \mbox{if $T+\tau_p \leq t \leq T+2\tau_p$}\,, 
\end{array} \right.
\end{equation}
with $\Omega_g \tau_p \approx \pi/2$ and $T \gg \tau_p$. 
The effect of the first pulse is to evolve the initial state into a coherent
superposition of the initial and final states.  During the delay time between
pulses, the system carries out phase oscillations.  Finally, the second pulse 
rotates the state vector again by an angle of $\Omega_g\tau_p$.  Fixing the
delay time $T$ and measuring the population in the final state as a function of the 
detuning $\delta$ at the final time $T+2\tau_p$ yields a fringe pattern as shown in
Fig.~\ref{Fig_Ramsey_Fringe_QR}. The figure shows the Ramsey fringes obtained
for a single QR and the value of $\Delta_{QR}$ such that $\delta = 0$ is easy to
identify from the fringe pattern.

\section{Magnetometer}  \label{Sec:Magnetometer}

Atomic magnetometers often rely on a measurement of the Larmor precession of spin-polarized
atoms in a magnetic field \cite{Budker_07}.  One of the limitations
on their sensitivity is spin relaxation.  In our system, spin relaxation
is highly suppressed if the lattice is singly occupied.

The degenerate ground state of a fermionic QR is split by the external field,
and measuring the frequency splitting can accurately determine
the external field.
When the QR is placed in an external magnetic field, say
$B_{\mathrm{ex}} \mbfe_z$ (for simplicity), a Zeeman interaction term,
$H_B = -g \mu_B B_{\mathrm{ex}} F_z$, must be included in the
Schr\"odinger equation (\ref{eq-Schrodinger-def}).
%
The energy of the ground state calculated to first order
in $B_{\mathrm{ex}}$ is
\begin{eqnarray}
  \epsilon_{0,\zeta}(B_{\mathrm{ex}}) &=&
  \epsilon_{0,\zeta} +
  \frac{2 \zeta g \mu_B B_{\mathrm{ex}}}{2 I+1}~
  \beta_{0,1/2}^{z}.
  \label{energy-vs-Bz}
\end{eqnarray}
With $V_0=100 \, {\mathcal{E}}_{0}$ and $B_0=180 \, {\mathcal{E}}_{0}$,
$\beta_{0,1/2}^{z}=0.1078$. 
%
Equation (\ref{energy-vs-Bz}) shows that the external magnetic field
splits the degeneracy of the energy levels with $\zeta=\pm 1/2$.
The Raman scattering between these levels gives rise to Rabi oscillations
with amplitude that has a maximum when $\omega_p-\omega_s=\Delta_B$,
where
\begin{eqnarray}
  \Delta_B  &=&
  \frac{{g \mu_B B_{\mathrm{ex}}}}{(2 I + 1) \hbar}~
  \beta_{0,1/2}^{z}.
  \label{frequency-difference}
\end{eqnarray}


The frequency splitting $\Delta_B$ can be experimentally measured,
therefore the QR can be used as a magnetometer: measuring $\Delta_B$
and comparing with Eq.~(\ref{frequency-difference}) yields the external 
magnetic field $B_{\mathrm{ex}}$. The uncertainty of $B_{\mathrm{ex}}$ 
results largely from the uncertainty of $\beta_{0,1/2}^{(z)}$ which is 
a function of the laser frequency $\omega$ and amplitude $E_0$.
Hence, the accuracy of measuring $B_{\mathrm{ex}}$ is
\begin{eqnarray}
  \frac{\delta B_{\mathrm{ex}}}
       {B_{\mathrm{ex}}} =
  \frac{1}{\beta_{0,1/2}^{z}}~
  \sqrt{\big(
             \frac{\partial \beta_{0,1/2}^{z}}
                  {\partial \omega}~
             \delta \omega
        \big)^{2}+
        \big(
             \frac{\partial \beta_{0,1/2}^{z}}
                  {\partial E_0}~
             \delta E
        \big)^{2}},
\label{accuracy-Bz-def}
\end{eqnarray}
where $\delta \omega$ and $\delta E$ are the uncertainties
of $\omega$ and $E_0$.
Additional analysis of $\delta B_{\mathrm{ex}}$,
including the suppression of spin relaxation in a singly occupied
optical lattice \cite{Viverit_04, Bloch_08}, directional external magnetic
field effects, the lack of spectral line splitting due to the small 
anisotropy of the effective lattice potential, and measurement-time 
limitations, is provided in Sec.~\ref{Sec:Accuracy}.

\section{Rotation Sensor}  \label{Sec:Rotation Sensor}

When the QR is in a non-inertial frame rotating with
an angular velocity $\boldsymbol\Omega$ there is
an additional term $H_{\Omega} = \hbar {\boldsymbol \Omega} \cdot
{\boldsymbol {\mathcal{L}}}$ in the Hamiltonian,
where ${\boldsymbol {\mathcal{L}}}=\mbfF + \boldsymbol\ell$.
For $\boldsymbol \Omega$ along the $z$-axis, the QR energy is,
\begin{eqnarray}
  \epsilon_{0,\zeta}(\Omega_z) &=&
  \epsilon_{0,\zeta} +
  \zeta \hbar \Omega_z .
  \label{energy-vs-Omega}
\end{eqnarray}
In fact, for arbitrary $\boldsymbol \Omega$, Eq.~(\ref{energy-vs-Omega}) is valid 
to first order in $\boldsymbol\Omega$ (see Appendix \ref{Sec:Distinguishing} for details).
The splitting of the ground state energy is 
$\Delta_{\Omega} = \Omega_z$, and the accuracy of measurement of the angular velocity
due to spontaneous magnetic dipole transitions within the ground state QR manifold is
$\delta \Omega_z = \delta \Delta_{\Omega} = 4 g^2 \mu_{B}^{2} \Omega_{z}^{3}/(3 \hbar c^3)$.
Additional analysis of $\delta \Omega_z$,
including the suppression of spin relaxation in a singly occupied
optical lattice \cite{Viverit_04, Bloch_08},
the lack of spectral line splitting due to the small 
anisotropy of the effective lattice potential, and measurement-time 
limitations, is given in Sec.~\ref{Sec:Accuracy}.

\section{Accelerometer}  \label{Sec:Accelerometer}

When the QR is in a non-inertial frame moving with
a linear acceleration $\mbfa$, an additional term $H_a = M {\bf a} \cdot {\bf r}$ must be
included in the Hamiltonian. The energy of the ground state calculated to first order 
in $\mbfa$ is
\begin{eqnarray}
  \epsilon_{0,\zeta}(a_{\parallel}) &=&
  \epsilon_{0,\zeta} +
  \zeta M a_{\parallel} \varrho_{0,\zeta},
  \label{energy-vs-a}
\end{eqnarray}
where $\varrho_{n,\zeta} = \int[\psi_{n,\zeta,1/2}^{2}(r) + \psi_{n,\zeta,-1/2}^{2}(r)] r dr$.
The splitting of the ground state energy is $\Delta_a = M a_{\parallel} \varrho_{0,\zeta}/\hbar$.
The acceleration measurement accuracy $\delta a_{\parallel}$ is considered in
Sec.~\ref{Sec:Accuracy}.

\section{Accuracy Estimates for ${\bf B}_{\mathrm{ex}}$, ${\boldsymbol \Omega}$, ${\bf a}$} 
\label{Sec:Accuracy}

$ \delta {\bf B}_{\mathrm{ex}}$:  
The uncertainty of $B_{\mathrm{ex}}$ results largely from the uncertainty
of the quantity $\beta_{0,1/2}^{z}$ which is a function of the laser frequency $\omega$
and amplitude $E_0$. Hence, the accuracy of measuring $B_{\mathrm{ex}}$
is given by Eq.~(\ref{accuracy-Bz-def}).
It is convenient to rewrite Eq.~(\ref{accuracy-Bz-def}) as
\begin{eqnarray}
  \frac{\delta B_{\mathrm{ex}}}
       {B_{\mathrm{ex}}} =
  \frac{1}{\beta_{0,1/2}^{z}}~
  \sqrt{\big(
             \frac{\partial \beta_{0,1/2}^{z}}
                  {\partial {\mathcal{E}}_{0}}~
             \delta {\mathcal{E}}
        \big)^{2}+
        \big(
             \frac{\partial \beta_{0,1/2}^{z}}
                  {\partial {\mathcal{I}}_{0}}~
             \delta {\mathcal{I}}
        \big)^{2}},
  \label{accuracy-Bz-recoil-intensity}
\end{eqnarray}
where $\delta {\mathcal{E}}$ and $\delta {\mathcal{I}}$
are uncertainties of the recoil energy ${\mathcal{E}}_{0}$
and the laser intensity ${\mathcal{I}}_{0}$.
$\beta_{0,1/2}^{z}$ depends on a single parameter, 
${\mathcal{I}}_{0}/{\mathcal{E}}_0$, therefore
$$
  {\mathcal{E}}_{0}~
  \frac{\partial \beta_{0,1/2}^{z}}
       {\partial {\mathcal{E}}_{0}}
  ~=~
  -{\mathcal{I}}_{0}~
  \frac{\partial \beta_{0,1/2}^{z}}
       {\partial {\mathcal{I}}_{0}}.
$$
Assuming that $\delta {\mathcal{I}}/{\mathcal{I}}_{0} \ll \delta \omega/\omega$,
we get
$$
  \frac{\delta B_{\mathrm{ex}}}
       {B_{\mathrm{ex}}}
  =
  \frac{\delta {\mathcal{E}}}{\beta_{0,1/2}^{z}}~
  \bigg|
       \frac{\partial \beta_{0,1/2}^{z}}
            {\partial {\mathcal{E}}_{0}}
  \bigg|.
$$

Numerical calculations for $V_0=100 \, {\mathcal{E}}_{0}$
and $B_0=180 \, {\mathcal{E}}_{0}$ give
$$
  \frac{\delta B_{\mathrm{ex}}}
       {B_{\mathrm{ex}}}
  ~=~
  -\frac{0.075495~\delta \omega}
        {\omega},
$$
where we have used the fact that
$\delta {\mathcal{E}}/{\mathcal{E}}_{0} = 2 \, \delta \omega/\omega$.
For lithium atoms, $\omega = 2.808 \times 10^{15}$~s$^{-1}$
\cite{Li-energy-PRA-1995}, and taking $\delta \omega = 2 \pi \times 160$~mHz~$= 1.005$~s$^{-1}$
\cite{mHz-laser-OptLett-2014}, we find
\begin{eqnarray}
  \frac{\delta B_{\mathrm{ex}}}
       {B_{\mathrm{ex}}}
  &=&
  2.703 \times 10^{-17}.
  \label{deltaB-to-B}
\end{eqnarray}

$\delta{\bm \Omega}_{\mathrm{z}}$:  
As already stated in Sec.~\ref{Sec:Rotation Sensor}, the accuracy of measurement 
of the angular velocity due to
spontaneous magnetic dipole transitions within the ground
state QR manifold is
$\delta \Omega_z = 4 g^2 \mu_{B}^{2} \Omega_{z}^{3}/(3 \hbar c^3)$,
hence
\begin{equation} \label{delta-Omega-res}
  \frac{\delta \Omega_z}{\Omega_{z}^{3}}
  = 1.614 \times 10^{-44}~{\text{s}}^2.
\end{equation}
For $\Omega_z=72.722~\mu$rad/s (the rotation frequency of Earth),
$\delta \Omega_z = 6.209 \times 10^{-57}$~s$^{-1}$.

{$\delta{\bf a}$:
The acceleration measurement accuracy $\delta a_{\parallel}$ is given by
\begin{equation} \label{delta-a_parallel}
  \frac{\delta a_{\parallel}}
       {a_{\parallel}} =
  \frac{1}{\varrho_{0,1/2}}~
  \sqrt{\Big(
             \frac{\partial \varrho_{0,1/2}}
                  {\partial \omega}~
             \delta \omega
        \Big)^{2}+
        \Big(
             \frac{\partial \varrho_{0,1/2}}
                  {\partial E_0}~
             \delta E
        \Big)^{2}}.
\end{equation}
Taking into account the equalities,
\begin{eqnarray*}
  \frac{\partial \varrho_{0,1/2}}{\partial \omega}~
  \delta \omega
  &=&
  \frac{\partial \varrho_{0,1/2}}{\partial {\mathcal{E}}_{0}}~
  \delta {\mathcal{E}},
  \\
  \frac{\partial \varrho_{0,1/2}}{\partial E_0}~
  \delta E
  &=&
  \frac{\partial \varrho_{0,1/2}}{\partial {\mathcal{I}}_{0}}~
  \delta {\mathcal{I}},
\end{eqnarray*}
where ${\mathcal{E}}_{0}$ is the recoil energy (\ref{recoil-energy-def})
and ${\mathcal{I}}_{0} \propto E_{0}^{2}$ is the intensity of the laser beam.
Assuming that
$$
  \bigg|
       \frac{\partial \varrho_{0,1/2}}{\partial {\mathcal{I}}_{0}}
  \bigg|~
  \delta {\mathcal{I}}
  ~\ll~
  \bigg|
       \frac{\partial \varrho_{0,1/2}}{\partial {\mathcal{E}}_{0}}
  \bigg|~
  \delta {\mathcal{E}},
$$
we find that
\begin{eqnarray*}
  \frac{\delta a_{\parallel}}
       {a_{\parallel}} \approx
  \frac{\delta {\mathcal{E}}}{\varrho_{0,1/2}}~
  \bigg|
       \frac{\partial \varrho_{0,1/2}}
            {\partial {\mathcal{E}}_{0}}
  \bigg|.
\end{eqnarray*}
Numerical calculations for $V_0=100 \, {\mathcal{E}}_{0}$
and $B_0=180 \, {\mathcal{E}}_{0}$ give
$$
  \frac{1}{\varrho_{0,1/2}}~
  \bigg|
       \frac{\partial \varrho_{0,1/2}}
            {\partial {\mathcal{E}}_{0}}
  \bigg|
  =
  \frac{0.139986}{{\mathcal{E}}_{0}},
$$
hence
$$
  \frac{\delta a_{\parallel}}
       {a_{\parallel}} =
  1.002 \times 10^{-16}.
$$

The measurement accuracy of an external magnetic field, angular velocity 
and linear acceleration are also affected by the energy-time uncertainty 
principle.  Given a measurement time $T$, the measurement bandwidth is $1/(2T)$, 
and the accuracy of the QR magnetometer, accelerometer or gyroscope is proportional
to $1/\sqrt{T}$ \cite{Budker-Kimball-book}.
For an optical lattice with $N$ QRs, the accuracies of the magnetometer, 
gyroscope and accelerometer are \cite{Budker-Kimball-book}
\begin{eqnarray}
  &&
  \delta B_u =
  \frac{B_{\mathrm{ex}}}{\beta_{0,1/2}^{z}}~
  \bigg|
       \frac{\partial \beta_{0,1/2}^{z}}
            {\partial \omega}
  \bigg|~
  \sqrt{\frac{\delta \omega}{N T}}, \\
  && 
  \delta \Omega_u =
  \sqrt{\frac{\delta \Omega_z}{N T}}, \\
  &&
  \delta a_u =
  \frac{a_{\parallel}}{\varrho_{0,1/2}}~
  \bigg|
       \frac{\partial \varrho_{0,1/2}}
            {\partial \omega}~
  \bigg|~
  \sqrt{\frac{\delta \omega}{N T}}.
\end{eqnarray}
Here the subscript $u$ denotes uncertainty due to the finite measurement time,
$\delta \omega$ is the uncertainty of the optical lattice laser frequency $\omega$,
and $\delta \Omega_z$ is the angular velocity uncertainty given by 
Eq.~(\ref{delta-Omega-res}).  Taking $T=1$~s, and $N=4.99 \times 10^8$ (which 
corresponds to an optical lattice of area 1~cm$^2$), we get
\begin{eqnarray}
  \frac{\delta B_u}{B_{\mathrm{ex}}}
  &=&
  1.74 \times 10^{-20},
  \\
  \frac{\delta \Omega_u}{\Omega_{z}^{3/2}}
  &=&
  5.687 \times 10^{-27}~{\text{s}}^{1/2},
  \\
  \frac{\delta a_u}{a_{\parallel}}
  &=&
  4.474 \times 10^{-21}.
\end{eqnarray}

\section{Uncertainty due to Shot noise in the Stokes and pump pulses} \label{Sec:shot-noise}

Another source of uncertainty $\delta B_{\mathrm{ex}}$, $\delta \Omega$ and $\delta a$ arises
from shot noise in the Stokes and pump pulses used to measure the detuning $\Delta_{\mathrm{QR}}$
of the QR using the Ramsey separated field method (see Figs.~\ref{Fig-Raman-Rabi} and 
\ref{Fig_Ramsey_Fringe_QR}).  Shot noise results in fluctuations in the position and amplitude of the population oscillations of the Ramsey fringes
because the $\pi/2$ Raman pulses have Rabi frequencies which fluctuate $\Omega_R \tau$,
\begin{eqnarray*}
  \phi_R &\equiv& \Omega_R \tau =
  \frac{\pi}{2} \pm \delta \phi_R,
\end{eqnarray*}
where
\begin{eqnarray}
  \delta \phi_R
  ~\lesssim~
  \frac{\pi}{2}~
  \Bigg(
       \frac{1}{\sqrt{N_p}} +
       \frac{1}{\sqrt{N_s}}
  \Bigg),
  \label{delta-phi-shot-noise}
\end{eqnarray}
and $N_p$ and $N_s$ are the number of photons of the pump and Stokes beams, assuming a Poissonian
distribution of photon number.  The uncertainty $\delta \Delta_{\mathrm{QR}}$ of 
$\Delta_{\mathrm{QR}}$ can be estimated from the variation of the probability for population transfer
from the initial level of the split QR ground state level to the final level.  A simple calculation
shows that
\begin{eqnarray}
  \delta \Delta_{\mathrm{QR}} \lesssim
  \frac{\sqrt{1.5} \, \delta \phi_R}{\tau} \lesssim
  \frac{1.92}{\tau}~
  \Bigg(
       \frac{1}{\sqrt{N_p}} +
       \frac{1}{\sqrt{N_s}}
  \Bigg),
  \label{delta-QR}
\end{eqnarray}
where we have used (\ref{delta-phi-shot-noise}) for $\delta \phi_R$.

The ground-state hyperfine splitting of lithium atoms is
$\omega_{\mathrm{hf}} =
  2 \pi \times 228.2~{\text{MHz}} =
  1.434 \times 10^{9}~{\text{s}}^{-1}$,
so the energy of a Stokes and pump photons are
$\hbar \omega_s \lesssim \hbar \omega_p \lesssim \hbar \omega_{\mathrm{hf}}$,
where $\hbar \omega_{\mathrm{hf}} = 1.51226 \times 10^{-18}~{\text{erg}}$.
The numbers of photons $N_p$ and $N_s$ can be estimated as
\begin{eqnarray}
  N_{\nu} &=&
  \frac{\pi r_b^2 \tau I_{\nu}}{\hbar \omega_{\nu}},
  \label{number-of-photons}
\end{eqnarray}
where $I_{\nu}$ is the intensity of the pump ($\nu = p$) or Stokes ($\nu = s$) field,
$r_b$ is the radius of the beam (which is assumed to be the same for the pump and
Stokes beams) and $\tau$ is the pulse duration.
Taking $I_p = 5 \times 10^3$~W/m$^2$, $I_s = 2 \times 10^4$~W/m$^2$, we get
$\tau = 1.644$~ms. For $r_b \approx 0.2$~m (the microwave wavelength),
the numbers of photons (\ref{number-of-photons}) are
$$
  N_p \approx 7.467 \times 10^{24},
  \ \ \ \ \
  N_s \approx 2.987 \times 10^{25},
$$
so from Eq.~(\ref{delta-QR}) $\delta \Delta_{\mathrm{QR}}$ is
\begin{eqnarray*}
  \delta \Delta_{\mathrm{QR}} ~\approx~
  6.411 \times 10^{-10}~{\text{s}}^{-1} ~=~
  2 \pi \times 0.102~{\text{nHz}}.
\end{eqnarray*}

Knowing $\Delta_{\mathrm{QR}}$, the external magnetic field
$B_{\mathrm{ex}}$, the angular velocity $\Omega$ or the acceleration $a$
of the non-inertial frame can be calculated using
\begin{eqnarray}
  B_{\mathrm{ex}} &=&
  \frac{(2 I + 1) \hbar \Delta_{\mathrm{QR}}}
       {g \mu_B \beta^{z}_{0,1/2}},
  \label{B-vs-Delta}
  \\
  \Omega &=&
  \Delta_{\mathrm{QR}},
  \label{Omega}
  \\
  a &=&
  \frac{\hbar \Delta_{\mathrm{QR}}}
       {M \varrho_{0,1/2}}.
  \label{a-vs-Delta}
\end{eqnarray}
Here $I = 1$ is the nuclear spin of $^6$Li atoms,
$\beta_{0,1/2}^{z} = 0.1078$,
$M$ is the atomic mass and $\varrho_{0,1/2} = 0.0987 \lambda_0$, 
where $\lambda_0$ is the wavelength of the laser beam
creating the optical lattice. For $^6$Li atoms,
$\lambda_0 = 670.8$~nm, and therefore $\varrho_{0,1/2} = 66.18$~nm.

When the optical lattice has $N$ quantum rotors,
uncertainties of external magnetic field $B_{\mathrm{ex}}$,
angular velocity $\Omega$ and acceleration $a$ due to
the shot noise are given by,
\begin{eqnarray}
  \delta B_{\mathrm{ex}} &=&
  \frac{1}{\sqrt{N}}~
  \frac{(2 I + 1) \hbar~\delta \Delta_{\mathrm{QR}}}
       {g \mu_B \beta^{z}_{0,1/2}},
  \label{delta-B-shot-noise}
  \\
  \delta \Omega &=&
  \frac{\delta  \Delta_{\mathrm{QR}}}{\sqrt{N}},
  \label{delta-Omega-shot-noise}
  \\
  \delta a &=&
  \frac{1}{\sqrt{N}}~
  \frac{\hbar~\delta  \Delta_{\mathrm{QR}}}{M \varrho_{0,1/2}}.
  \label{delta-a-shot-noise}
\end{eqnarray}
For $N = 4.99 \times 10^8$ $^6$Li atoms,
\begin{eqnarray*}
  \delta B_{\mathrm{ex}} &=&
  4.54 \times 10^{-24}~{\text{Tesla}},
  \\
  \delta \Omega &=&
  2.87 \times 10^{-14}~{\text{s}}^{-1},
  \\
  \delta a &=&
  4.58 \times 10^{-15}~{\text{m/s}}^{2}.
\end{eqnarray*}
Compare this result with the uncertainties of the magnetic field and angular
velocity of the Earth, and the acceleration due to gravity derived in
Sec.~\ref{Sec:Accuracy},
\begin{eqnarray*}
  \delta B_u &=&
  7.8 \times 10^{-25}~{\text{Tesla}},
  \\
  \delta \Omega_u &=&
  3.6 \times 10^{-33}~{\text{s}}^{-1},
  \\
  \delta a_u &=&
  4.4 \times 10^{-20}~{\text{m/s}}^{2}.
\end{eqnarray*}
Hence the shot noise contribution to the uncertainty is larger than the 
uncertainty due to the decay rate of the excited hyperfine state.

\section{Summary and Conclusions}  \label{Sec:Summary}

The wave function of QRs, atoms trapped by a SDOLP, are confined to circular rings of 
radius $r_0$ with center at the minima of the scalar lattice potential.
SDOLPs with precisely one atom per site (for which suppress spin-relaxation) can be 
used as ultra-high accuracy rotation sensors, accelerometers or magnetometers.
The Ramsey time-separated oscillating field method with far-off-resonance Raman 
pulses between the split ground state of fermionic QRs can be used as a spectroscopic 
measurement technique for these applications, with a major accuracy limitation 
due to measurement-time uncertainty as outlined in Secs.~\ref{Sec:Accuracy} 
and \ref{Sec:shot-noise}.  Bosonic QRs have ground states which are
not degenerate, but their excited states are degenerate.  The splitting of the
excited states in the presence of rotation, in-plane acceleration or magnetic fields
can also be used for sensing.

\begin{acknowledgments}
This work was supported in part by a grant from the DFG through the DIP program (FO703/2-1).
\end{acknowledgments}

\appendix

\section{Optical lattice potential} \label{Sec:OLP}

Consider a 2D hexagonal optical lattice potential produced by six coherent 
laser beams having a superposition of in-plane and out-of-plane linear polarization with a 
configuration shown in Fig.~\ref{Fig-opt-latt-triangle}.  Note that other configurations, 
e.g., three or four laser beams instead of six, can also produce a SDOLP, but we shall not
consider other configurations here.  The wavelength, wave number, and frequency of the 
laser beams are $\lambda_0$, $q_0 = 2 \pi/\lambda_0$ and $\omega_0 = q_0 c$, where $c$ 
is the speed of light.  The optical lattice frequency should be 
slightly detuned to the red of resonance.  For ${}^{6}$Li atoms, the resonant wavelength is 
$\lambda_{\mathrm{res}} = 670.98$~{nm}. 
The resultant electric field is given by $\mbfE(\mbfr,t) =  [ \mbfE(\mbfr)  e^{-i \omega_0 t}+
{\mathrm{c.c.}} ]/2$ where the spatial dependence of the field is
\begin{eqnarray}
  \mbfE(\mbfr) =
  E_0
  \sum_{n=1}^{6}
  \boldsymbol\xi_n
  e^{i \mbfq_n \mbfr}.
  \label{EF=E1+E2+E3+E4+E5+E6}
\end{eqnarray}
The wavevectors $\mbfq_n$ are
\begin{eqnarray}
  &&
  \mbfq_1 ~=~
  -\mbfq_4 ~=~
  -q_0~
  \mbfe_x,
  \nonumber
  \\
  &&
  \mbfq_2 ~=~
  -\mbfq_5 ~=~
  -\frac{q_0}{2}~
  \Big\{
      \mbfe_x+
      \sqrt{3}~
      \mbfe_y
  \Big\},
  \label{q1-q2-q3-q4-q5-q6-def}
  \\
  &&
  \mbfq_3 ~=~
  -\mbfq_6 ~=~
  \frac{q_0}{2}~
  \Big\{
      \mbfe_x-
      \sqrt{3}~
      \mbfe_y
  \Big\},
  \nonumber
\end{eqnarray}
and the unit vectors $\mbfe_x$, $\mbfe_y$ and $\mbfe_z$ 
are parallel to the $x$, $y$ and $z$ axes.
The polarization vectors $\boldsymbol\xi_n$
($1 \leq n \leq 6$) are
$\boldsymbol\xi_n =
  \Big\{
      \sqrt{1-\beta^2}~
      \mbfe_z+
      \frac{\beta}{q_0}~
      \big[
          \mbfq_n
          \times
          \mbfe_z
      \big]
  \Big\}$,
where $\beta$ is real and lies
in the interval $0 \, < \, \beta \, \leq \, 1/\sqrt{2}$.
Hereafter, we take $\beta=1/\sqrt{2}$.

For an arbitrary atom with electronic angular momentum ${\bf J}$, 
the electric field generates an effective SDOLP \cite{SOI-EuroPhysJ-13},
\begin{eqnarray}
  U(\mbfr) &=&
  -\frac{1}{4}~
  \Bigg\{
       \alpha_{n,J}^{s}(\omega_0)~
       \mbfE^{*}(\mbfr)
       \cdot
       \mbfE(\mbfr)-
  \nonumber \\ && -
       \frac{i \alpha_{n,J}^{v}(\omega_0)}
            {2 J}~
       \big[
           \mbfE^{*}(\mbfr)
           \times
           \mbfE(\mbfr)
       \big]
       \cdot
       {\mathbf{J}}+
  \nonumber \\ && +
       \frac{3 \alpha_{n,J}^{t}(\omega_0)}
            {2 J (2 J-1)}~
       \Big[
           \big(
               \mbfE^{*}(\mbfr)
               \cdot
               {\mathbf{J}}
           \big)~
           \big(
               \mbfE(\mbfr)
               \cdot
               {\mathbf{J}}
           \big)+
  \nonumber \\ && ~~~ +
           \big(
               \mbfE(\mbfr)
               \cdot
               {\mathbf{J}}
           \big)~
           \big(
               \mbfE^{*}(\mbfr)
               \cdot
               {\mathbf{J}}
           \big)-
  \nonumber \\ && ~~~ -
           \frac{2}{3}~
           J (J+1)~
           \mbfE^{*}(\mbfr)
           \cdot
           \mbfE(\mbfr)
       \Big]
  \Bigg\},
  \label{U=Us+Uv+Ut}
\end{eqnarray}
where $U$ is a $(2J+1)$$\times$$(2J+1)$ matrix in spin-space.
Here $\alpha_{n,J}^{s}(\omega_0)$, $\alpha_{n,J}^{v}(\omega_0)$
and $\alpha_{n,J}^{t}(\omega_0)$ are the conventional dynamical
scalar, vector and tensor polarizabilities of the atom in the 
fine-structure level $|nJ\rangle$ with principal quantum number 
$n$ and total electronic angular momentum $J$ \cite{SOI-EuroPhysJ-13}:
\begin{eqnarray}
  \alpha_{n,J}^{s}(\omega_0) &=&
  \frac{\alpha_{n,J}^{(0)}(\omega_0)}
       {\sqrt{3 (2 J+1)}},
  \label{alpha-s-J}
  \\
  \alpha_{n,J}^{v}(\omega_0) &=&
  -\frac{\sqrt{2 J}~
         \alpha_{n,J}^{(1)}(\omega_0)}
        {\sqrt{(J+1)(2 J+1)}},
  \label{alpha-v-J}
  \\
  \alpha_{n,J}^{t}(\omega_0) &=&
  -\frac{\sqrt{2 J (2 J-1)}~
        \alpha_{n,J}^{(2)}(\omega_0)}
       {\sqrt{3 (J+1) (2 J+1) (2 J+3)}}.
  \label{alpha-t-J}
\end{eqnarray}
The terms on the right hand side of Eq.~(\ref{U=Us+Uv+Ut})
proportional to $\alpha_{J}^{s}(\omega_0)$,
$\alpha_{J}^{v}(\omega_0)$ and $\alpha_{J}^{t}(\omega_0)$
describe a spin-independent optical lattice potential,
a Zeeman-type interaction and a tensor Stark-type interaction
respectively.
The  scalar $\alpha_{n,J}^{(0)}(\omega_0)$,
vector $\alpha_{n,J}^{(1)}(\omega_0)$ and tensor
$\alpha_{n,J}^{(2)}(\omega_0)$ polarizabilities of
the atom in the fine-structure level $|n J\rangle$ can be
calculated as follows \cite{SOI-EuroPhysJ-13}:
\begin{eqnarray}
  \alpha_{n,J}^{(K)}(\omega_0) &=&
  \big(-1\big)^{K+J+1}
  \sqrt{2 K+1}
  \sum_{n',J'}
  \big(-1\big)^{J'}
  \times \nonumber \\ && \times
  \left\{
    \begin{array}{ccc}
      1 & K & 1
      \\
      J & J' & J
    \end{array}
  \right\}~
  \Big|
      \big\langle
          n'J'
      \big\|
          \mbfd
      \big\|
          nJ
      \big\rangle
  \Big|^{2}
  \times \nonumber \\ && \times
  \frac{1}{\hbar}~
  {\mathrm{Re}}
  \bigg(
      \frac{1}{\omega_{n',J';n,J}-\omega_0-i\gamma_{n',J';n,J}/2}+
  \nonumber \\ && ~~ +
      \frac{(-1)^{K}}{\omega_{n',J';n,J}+\omega_0+i\gamma_{n',J';n,J}/2}
  \bigg).
  \label{polar-reduced-def}
\end{eqnarray}
Here $K = 0, 1,2$ gives the scalar, vector and tensor polarizibilities, 
$\Big\{\begin{array}{ccc}1 & K & 1 \\ J & J' & J \end{array}\Big\}$
is the Wigner 6-$j$ symbol, and $\langle{nJ}\parallel\mbfd\parallel{n'J'}\rangle$ is 
the reduced matrix elements of the dipole moment operator.
The quantities $\omega_{n',J';n,J}=(\epsilon_{n',J'}-\epsilon_{n,J})/\hbar$
and $\gamma_{n',J';n,J}=\gamma_{n',J'}+\gamma_{n,J}$
are the angular frequency and linewidth of the transition between the 
fine-structure levels $|n J\rangle$ and $|n' J'\rangle$.

\subsection{Li atom scalar and vector polarizabilities}

Equation (\ref{polar-reduced-def}) contains an expression for the 
scalar and vector polarizabilities of an atom in terms of 
matrix elements of the dipole operator between electronic wave functions.  
Here we use this expression in the special case of $J=1/2$ and compute
the scalar and vector polarizabilities $\alpha_s(\omega)$ and
$\alpha_v(\omega)$ of the Li atom in its ground-state 
whose electronic configuration (outside the closed 1s shell) is
$2s~{}^2$S$_{1/2}$ \cite{Li-polarizability-2017},
\begin{subequations}   \label{subeqs-alpha-s-v}
\begin{eqnarray}
  \alpha_0(\omega) &=&
  -\frac{\alpha_{2,1/2}^{(0)}(\omega)}{\sqrt{6}},
  \label{alpha-s-def}
\end{eqnarray}
\begin{eqnarray}
  \alpha_1(\omega) &=&
  \frac{\alpha_{2,1/2}^{(1)}(\omega)}{\sqrt{3}},
  \label{alpha-v-def}
\end{eqnarray}
\end{subequations}
where $\alpha_{n,J}^{(K)}(\omega)$ ($K=0,1$) are the
coupled polarizabilities given in Eq.~(\ref{polar-reduced-def}),
whereas the tensor polarizability $\alpha_{n,J}^{(K=2)}(\omega) = 0$,
see Ref.~\cite{Li-polarizability-2017} for details.
For the calculations of the reduced matrix elements
$\la n'J' \vert \vert {\bf d} \vert \vert nJ \ra$ we need the
electronic wave function of  the Li atom in its ground and excited 
states. The ground state wave function with configuration $2s~{}^2$S$_{1/2}$ is
\begin{eqnarray}
  \Psi_{S,m_s}(r) &=&
  \frac{1}{\sqrt{4\pi}}~
  \psi_{2s}(r)
  \chi_{m_s},
  \label{WF-ground-state}
\end{eqnarray}
where $\psi_{2s}(r)$ is the real valued radial wave function of the $2s$
electron and $\chi_{m_s}$ are spin wave functions,
$m_s=\pm1/2$. The radial and spin wave functions are normalized by
the conditions,
$\int\limits_{0}^{\infty} \psi_{2s}^{2}(r) r^2 dr = 1$, and 
$\chi_{m_s}^{\dag} \chi_{m'_s} = \delta_{m_s,m'_s}$.

The electronic wave functions of Li in the excited $2p~{}^2$P$_{1/2}$ and
$2p~{}^2$P$_{3/2}$ states are given by
\begin{eqnarray}
  \Psi_{2p,J,m_J}(\mbfr) &=&
  \psi_{2p}(r)
  \sum_{m_L,m_s}
  C^{J,m_J}_{1,m_L;1/2,m_s}
  \times \nonumber \\ && \times
  Y_{1,m_L}(\theta,\phi)
  \chi_{m_s}.
  \label{WF-excited-state}
\end{eqnarray}
Here $\psi_{2p}(r)$ is the radial wave function of the $2p$ electron,
$Y_{1,m_L}(\theta,\phi)$ are spherical harmonics with $m_L=0,\pm1$, 
$\chi_{m_s}$ are spin wave functions with $m_s=\pm1/2$, and
$$
  C^{J,m_J}_{1,m_L;1/2,m_s}
  \equiv \Big\langle 1m_L~ \tfrac{1}{2} m_s
  \Big| J m_J \Big\rangle
$$
are Clebsch-Gordan coefficients.
The radial wave functions and the spherical harmonics are
normalized by the conditions, $\int\limits_{0}^{\infty}
\psi_{2p}^{2}(r) r^2 dr = 1$, and
\begin{eqnarray*}
  &&
  \int\limits_{0}^{2\pi} d\phi
  \int\limits_{0}^{\pi} \sin\theta d\theta~
  Y_{1,m_L}^{*}(\theta,\phi)
  Y_{1,m'_L}(\theta,\phi)
  = \delta_{m_L,m'_L}.
\end{eqnarray*}

The reduced matrix element of the electric dipole moment operator
is,
\begin{eqnarray}    \label{dipole-reduced}
  d_0
  \equiv
 \Big|
  \big\langle
       eJ
  \big\|
      \mbfd
  \big\|
       g 1/2
  \big\rangle
  \Big| 
  =
  e \int\limits_{0}^{\infty} \psi_{2s}(r)
  \psi_{2p}(r) r^3 dr .
\end{eqnarray}
The 6-j symbols required in Eq.~(\ref{polar-reduced-def}) for the Li atom are, 
$\Big\{\begin{array}{ccc}1 & K & 1 \\ 1/2 & J &  1/2  \end{array}\Big\}$, and
are given by
\begin{eqnarray*}
  \begin{array}{|c|c|c|}
    \hline
    &
    J = 1/2 &
    J = 3/2
    \\
    \hline
    K = 0 &
    1/\sqrt{6} &
    -1/\sqrt{6}
    \\
    \hline
    K = 1 &
    -1/3 &
    -1/6
    \\
    \hline
  \end{array}.
\end{eqnarray*}
We also need the resonant frequencies,
\begin{eqnarray*}
  \omega_{J,1/2} =
  \frac{1}{\hbar}~
  \big\{
      \epsilon_{{\mathrm{p}},J}-
      \epsilon_{{\mathrm{s}},1/2}
  \big\},
  \ \ \
  J = \frac{1}{2},~\frac{3}{2},
\end{eqnarray*}
where $\epsilon_{{\mathrm{s}},1/2}$ is the energy of the ground
$^2$S$_{1/2}$ state, and $\epsilon_{{\mathrm{p}},J}$ are
the energies of the excited $^2$P$_{J}$ states.  The optical
lattice frequency detuning from resonance should be smaller than
the fine structure splitting (to have a significant effective
magnetic field) but larger than the linewidth of the transitions.
Hence, we assume that the frequency $\omega$ satisfies the inequalities,\\
$$
  \frac{\gamma_{1/2,1/2}}{2}
  ~\ll~
  \omega_{1/2,1/2}-
  \omega
  ~\ll~
  \omega_{3/2,1/2}-
  \omega_{1/2,1/2},
$$
which imply
$$
  \frac{1}{\omega_{1/2,1/2}-\omega}
  ~\gg~
  \frac{1}{\omega_{3/2,1/2}-\omega}
  ~\gg~
  \frac{1}{\omega_{J,1/2}+\omega}.
$$
The last inequality shows that the main contribution to
the coupled polarizabilities (\ref{polar-reduced-def})
is from $(\omega_{J,1/2}-\omega)^{-1}$  (where $J=1/2,3/2$), 
whereas the terms $(\omega_{J,1/2}+\omega)^{-1}$ can be neglected.
As a result, the coupled polarizabilities (\ref{polar-reduced-def})
are given by,
\begin{eqnarray*}
  \alpha_{2,1/2}^{(0)}(\omega) &=&
  -\frac{d_{0}^{2}}{\sqrt{6} \hbar}~
  \frac{1}{\omega_{1/2,1/2}-\omega}-
  \frac{d_{0}^{2}}{\sqrt{6} \hbar}~
  \frac{1}{\omega_{3/2,1/2}-\omega},
  \\
  \alpha_{2,1/2}^{(1)}(\omega) &=&
  \frac{d_{0}^{2}}{\sqrt{3} \hbar}~
  \frac{1}{\omega_{1/2,1/2}-\omega}-
  \frac{d_{0}^{2}}{2 \sqrt{3} \hbar}~
  \frac{1}{\omega_{3/2,1/2}-\omega}.
\end{eqnarray*}
The scalar and vector polarizabilities (\ref{subeqs-alpha-s-v})
are,
\begin{subequations}    \label{subeqs-alpha-s-v-res}
\begin{eqnarray}
  \alpha_{0}(\omega) =
  \frac{d_{0}^{2}}{6 \hbar}~
  \frac{1}{\omega_{1/2,1/2}-\omega}+
  \frac{d_{0}^{2}}{6 \hbar}~
  \frac{1}{\omega_{3/2,1/2}-\omega},
  \label{alpha-s-res}
  \\
  \alpha_{1}(\omega) =
  \frac{d_{0}^{2}}{3 \hbar}~
  \frac{1}{\omega_{1/2,1/2}-\omega}-
  \frac{d_{0}^{2}}{6 \hbar}~
  \frac{1}{\omega_{3/2,1/2}-\omega}.
  \label{alpha-v-res}
\end{eqnarray}
\end{subequations}
The ratio of the  vector and scalar polarizabilities is,
\begin{eqnarray}
  \frac{\alpha_{1}(\omega)}
       {\alpha_{0}(\omega)}
  =
  \frac{2 \omega_{3/2,1/2}-\omega_{1/2,1/2}-\omega}
       {\omega_{3/2,1/2}+\omega_{1/2,1/2}-2 \omega}
  \approx
  2.
  \label{ratio-alpha-s-v}
\end{eqnarray}

\subsection{Spin-Dependent Optical Potential for $J=1/2$}
  \label{apend-OP-J=1/2}

The effective SDOLP $U({\bf r})$ in Eq.~(\ref{U=Us+Uv+Ut}) acts 
in the Hilbert space of atomic states spanned 
by the basis kets $\vert nJFf \rangle$ where the quantum number $F$ 
corresponds to the total atomic angular momentum operator ${\bf F} = {\bf J} + {\bf I}$
and $f$ is its projection on the $z$-axis \cite{SOI-EuroPhysJ-13}.
Generically, $U({\bf{r}})$ is not diagonal in $F$ nor in $f$.  However,   
when the off-diagonal elements with $F \ne F'$ are much smaller than the hyperfine 
energy splitting  of the atoms, we can neglect them. Within the subspace 
of fixed $n,J,F$  the matrix elements of $U(\bfr)$ are written as,
\begin{eqnarray*}
  W_{f,f'}(\mbfr) &=&
  \big\langle
      nJFf
  \big| U(\mbfr) \big|
      nJFf'
  \big\rangle.
\end{eqnarray*}

As already stated, in the special case $J=1/2$, $\alpha_{n,J}^{(K=2)}(\omega_0)=0$, 
i.e., the tensor Stark-type interaction operator vanishes. Hence, the 
optical lattice potential $W_{f,f'}(\mbfr)$ takes the form \cite{SOI-EuroPhysJ-13},
\begin{eqnarray}
  W_{f,f'}(\mbfr) &=&
  V(\mbfr)
  \delta_{f,f'}-
  \mbfB(\mbfr)
  \cdot
  \mbfF_{f,f'},
  \label{W=V+BF}
\end{eqnarray}
where the scalar optical potential $V(\mbfr)$
and an {\em fictitious magnetic field} $\mbfB(\mbfr)$
(which is taken to have units of energy) are
given by Eq.~(\ref{subeqs-V-B-def}).
The scalar and vector dynamical polarizabilities of the atom
are \cite{SOI-EuroPhysJ-13},
\begin{eqnarray}
  \alpha_0(\omega_0) &=&
  \frac{\alpha_{n,1/2}^{(0)}(\omega_0)}{\sqrt{6}},
  \label{polar-scalar-res}
  \\
  \alpha_1(\omega_0) &=&
  \big(
      -1
  \big)^{I+1/2+F}~
  \frac{\alpha_{n,1/2}^{(1)}(\omega_0)}
       {\sqrt{3}},
  \label{polar-vector-res}
\end{eqnarray}

Substituting Eq.~(\ref{EF=E1+E2+E3+E4+E5+E6})
into Eq.~(\ref{V-def}), and using polar coordinates $\mbfr=(r,\phi)$, 
$x = r \cos\phi$, $y = r \sin\phi$, we obtain
\begin{eqnarray}
  V(\mbfr) &=&
  -\frac{V_0}{3}-
  \frac{V_0}{6}
  \sum_{m=0}^{2}
  \cos
  \bigg(
       \frac{2 \pi r}{\lambda_0}~
       \cos\big(\phi-m \theta_0\big)
  \bigg)-
  \nonumber \\ && -
  \frac{V_0}{18}
  \sum_{m=0}^{2}
  \cos
  \bigg(
       \frac{2 \sqrt{3} \pi r}{\lambda_0}~
       \sin\big(\phi-m \theta_0\big)
  \bigg),
  \label{V-polar}
\end{eqnarray}
where $\theta_0=2\pi/3$.  The potential strength $V_0$ is given by,
\begin{eqnarray}
  &&
  V_0 =
  \frac{9 \alpha_0(\omega_0)}{2}~
  E_{0}^{2}.
  \label{V0-a0-triangle-def}
\end{eqnarray}

Substituting Eq.~(\ref{EF=E1+E2+E3+E4+E5+E6})
into Eq.~(\ref{B-def}), we get
\begin{eqnarray}
  \mbfB(\mbfr) &=&
  B_r(\mbfr)
  \mbfe_r+
  B_{\phi}(\mbfr)
  \mbfe_{\phi}.
  \label{B-polar}
\end{eqnarray}
The components $B_r(\mbfr)$ and
$B_{\phi}(\mbfr)$ of $\mbfB(\mbfr)$
are expanded as,  
\begin{subequations}    \label{subeqs-Br-Bphi-polar}
\begin{eqnarray}
  B_r(\mbfr) &=&
  \frac{B_0}{9(2I+1)}
  \sum_{m=0}^{3}
  \cos\big(\phi+m\theta_0\big)
  \times \nonumber \\ && \times
  \sin
  \bigg(
       \frac{2 \pi r}{\lambda_0}~
       \cos\big(\phi-m\theta_0\big)
  \bigg)+
  \nonumber \\ &+&
  \frac{B_0}{3 \sqrt{3} (2I+1)}
  \sum_{m=0}^{2}
  \sin\big(\phi-m\theta_0\big)
  \times \nonumber \\ && \times
  \sin
  \bigg(
       \frac{2 \sqrt{3} \pi r}{\lambda_0}~
       \sin\big(\phi-m\theta_0\big)
  \bigg)+
  \nonumber \\ &+&
  \frac{B_0}{9 (2I+1)}
  \sum_{m=0}^{2}
  \cos\big(\phi-m\theta_0\big)
  \times \nonumber \\ && \times
  \sin
  \bigg(
       \frac{4 \pi r}{\lambda_0}~
       \cos\big(\phi-m\theta_0\big)
  \bigg),
  \label{Br-polar}
\end{eqnarray}
\begin{eqnarray}
  B_{\phi}(\mbfr) &=&
  -\frac{B_0}{9 (2I+1)}
  \sum_{m=0}^{3}
  \sin\big(\phi+m\theta_0\big)
  \times \nonumber \\ && \times
  \sin
  \bigg(
       \frac{2 \pi r}{\lambda_0}~
       \cos\big(\phi-m\theta_0\big)
  \bigg)+
  \nonumber \\ &+&
  \frac{B_0}{3 \sqrt{3} (2I+1)}
  \sum_{m=0}^{2}
  \cos\big(\phi-m\theta_0\big)
  \times \nonumber \\ && \times
  \sin
  \bigg(
       \frac{2 \sqrt{3} \pi r}{\lambda_0}~
       \sin\big(\phi-m\theta_0\big)
  \bigg)+
  \nonumber \\ &+&
  \frac{B_0}{9 (2I+1)}
  \sum_{m=0}^{2}
  \sin\big(\phi-m\theta_0\big)
  \times \nonumber \\ && \times
  \sin
  \bigg(
       \frac{4 \pi r}{\lambda_0}~
       \cos\big(\phi-m\theta_0\big)
  \bigg),
  \label{B-phi-polar}
\end{eqnarray}
\end{subequations}
where the fictitious magnetic field strength $B_0$ is,
\begin{eqnarray}
  B_0 &=&
  \frac{9 \alpha_1(\omega_0)}{2}~
  E_{0}^{2}.
  \label{B0-triangle-def}
\end{eqnarray}

The expansions~(\ref{V-polar}) and (\ref{subeqs-Br-Bphi-polar})
assure that $V(r,\phi)$, $B_r(r,\phi)$ and $B_{\phi}(r,\phi)$ are 
invariant under the transformation $\phi\to\phi'=\phi-m \pi/3$, where 
$m$ is an integer.  Thus, the optical lattice potential $V(\mbfr)$
and the fictitious magnetic field $\mbfB(\mbfr)$
are invariant under rotations by
$\pi/3$ radians around the $z$-axis (see Fig.~\ref{Fig-opt-latt-triangle}),
\begin{eqnarray}
  &&
  V
  \bigg(
       U\Big(\frac{m\pi}{3}\Big)\mbfr
  \bigg) =
  V(\mbfr),
  \nonumber
  \\
  &&
  U\Big(\frac{m\pi}{3}\Big)
  \mbfB
  \bigg(
       U\Big(\frac{m\pi}{3}\Big)\mbfr
  \bigg)
  =
  \mbfB(\mbfr),
  \label{rotation-symmatry}
\end{eqnarray}
where the rotation matrix $U(\phi)$ is
$$
  U(\phi) =
  \left(
    \begin{array}{cc}
      \cos\phi &
      -\sin\phi
      \\
      \sin\phi &
      \cos\phi
    \end{array}
  \right).
$$

The optical potential (\ref{V-polar})
and the fictitious magnetic field
(\ref{B-polar}) are shown in
Fig.~\ref{Fig-V-tr} and \ref{Fig-B-tr}
for $\beta=1/\sqrt{2}$.

\begin{figure} 
\centering
  \subfigure[]
  {\includegraphics[width = 0.9 \linewidth]
   {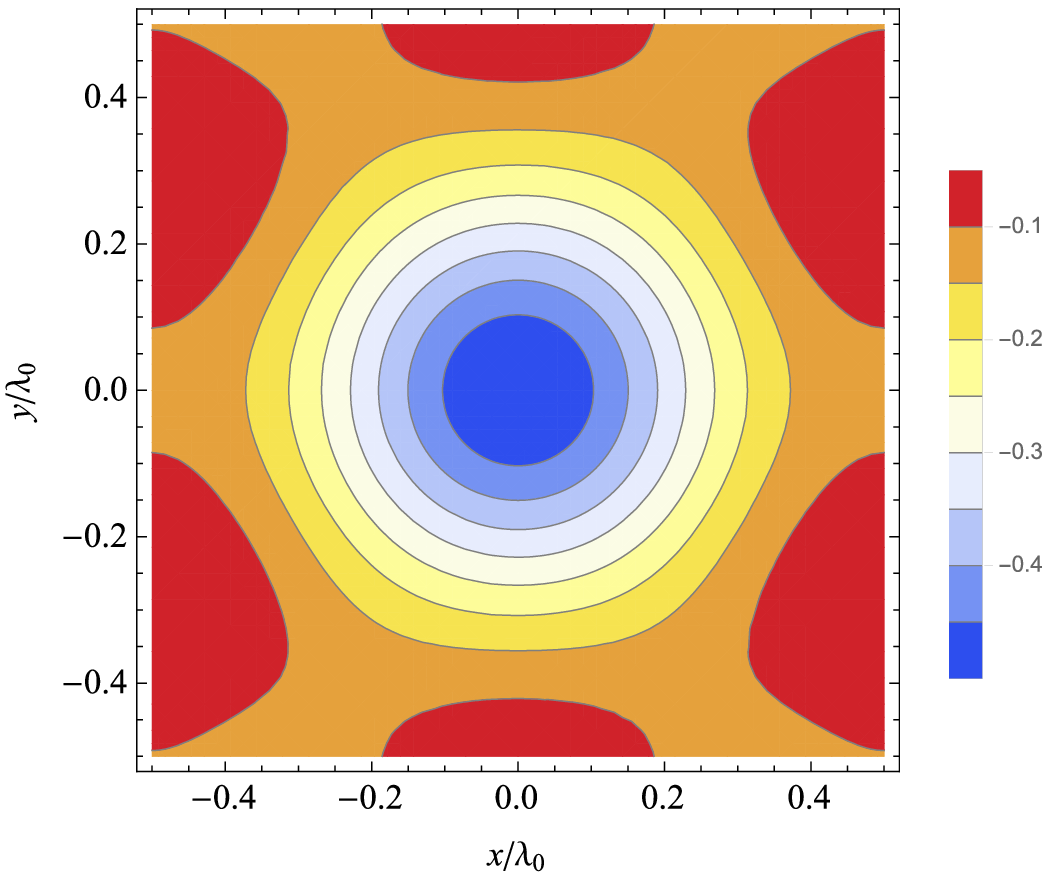}
   \label{Fig-V-tr}}
  \subfigure[]
  {\includegraphics[width = 0.9 \linewidth]
   {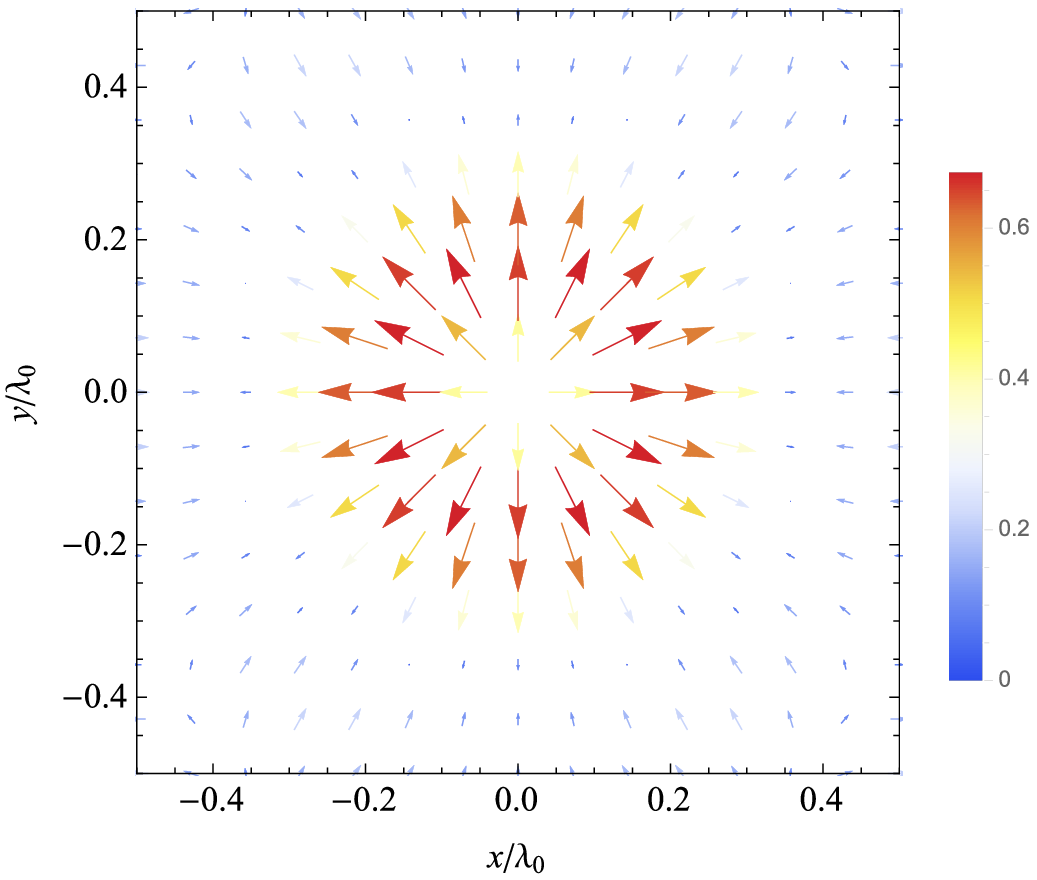}
   \label{Fig-B-tr}}
 \caption{
   (a) The optical lattice potential $V(\mbfr)$,
   Eq.~(\ref{V-polar}), and
   (b) the fictitious magnetic field
   (\ref{B-polar})
   as functions of the Cartesian
   coordinates $x$ and $y$ for
   $\beta=1/\sqrt{2}$.
   The plot legends denote
   (a) $V(\mbfr)/V_0$ (contour plot with 
   minimum indicated by deep blue).
   (b) $|\mbfB(\mbfr)|/B_0$
   (direction and strength indicated by arrows of varying length and color.
   $V_0$ and $B_0$ are given by
   Eqs.~(\ref{V0-a0-triangle-def}) and
   (\ref{B0-triangle-def}).}
 \label{Fig-V-B-tr}
\end{figure}


\section{Isotropic approximation for the spin-dependent optical lattice potential}
  \label{Sec:Isotropic}

The Fourier transform for $V(\mbfr)$,
$B_r(\mbfr)$ and $B_{\phi}(\mbfr)$ are
\begin{eqnarray}  \label{V_expansion}
  V(r,\phi) &=&
  \sum_{m}
  \tilde{V}_{m}(r)
  e^{6 i m \phi},
  \label{Fouries-V}
  \\
  B_r(r,\phi) &=&
  \sum_{m}
  \tilde{B}_{r,m}(r)
  e^{6 i m \phi},
  \label{Fourier-B-r}
  \\
  B_{\phi}(r,\phi) &=&
  \sum_{m}
  \tilde{B}_{\phi,m}(r)
  e^{6 i m \phi},
  \label{Fourier-B-phi}
\end{eqnarray}
where $m$ is an integer, and
$$
  \tilde{\mathcal{F}}_{m}(r) =
  \frac{1}{2\pi}
  \int\limits_{0}^{2\pi}
  {\mathcal{F}}(r,\phi)
  e^{-6 i m \phi} d\phi,
  \qquad
  {\mathcal{F}} = V, B_r, B_{\phi}.
$$
$V(r,\phi)$, $B_r(r,\phi)$ and $B_{\phi}(r,\phi)$
are real, and therefore
\begin{eqnarray*}
  \tilde{\mathcal{F}}_{-m}(r)
  &=&
  \tilde{\mathcal{F}}_{m}^{*}(r).
\end{eqnarray*}
Moreover, $V(r,\phi)$ and $B_r(r,\phi)$ are even
with respect to the inversion $\phi\to-\phi$,
and $B_{\phi}(r,\phi)$ is odd,
\begin{eqnarray*}
  V(r,-\phi) &=&
  V(r,\phi),
  \\
  B_r(r,-\phi) &=&
  B_r(r,\phi),
  \\
  B_{\phi}(r,-\phi) &=&
  -B_{\phi}(r,\phi).
\end{eqnarray*}
Therefore the Fourier components
$\tilde{V}_{m}$, $\tilde{B}_{r,m}$
and $\tilde{B}_{\phi,m}$ satisfy
the properties,
\begin{eqnarray*}
  \tilde{V}_{m}(r)
  &=&
  \tilde{V}_{-m}(r),
  \\
  \tilde{B}_{r,m}(r)
  &=&
  \tilde{B}_{r,-m}(r),
  \\
  \tilde{B}_{\phi,m}(r)
  &=&
  -\tilde{B}_{\phi,-m}(r).
\end{eqnarray*}

\subsection{Isotropic approximation for $V(\mbfr)$}

\begin{figure}
\centering
  \includegraphics[width=\linewidth,angle=0]
   {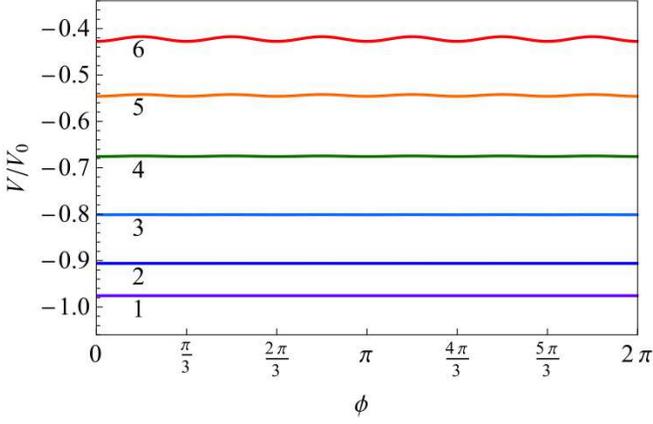}
 \caption{
   $V(\mbfr)$ (\ref{V-polar})
   as a function of the polar angle
   $\phi$ for $\beta=\frac{1}{\sqrt{2}}$.
   The purple (1), blue (2), sky blue (3),
   green (4), orange (5) and red (6)
   curves denote $r=0.05 \lambda_0$,
   $0.1 \lambda_0$, $0.15 \lambda_0$,
   $0.2 \lambda_0$, $0.25 \lambda_0$
   and $0.3 \lambda_0$.}
 \label{Fig-V-isotropic}
\end{figure}

Fig.~\ref{Fig-V-isotropic} shows $V(\mbfr)$ as a function of
$\phi$ for $\beta=1/\sqrt{2}$ and a few values of $r$.
It is clear that for $r<0.3 \lambda_0$, the potential is almost
isotropic.  Hence, for $r \lesssim 0.3 \, \lambda_0$, the optical potential
(\ref{V-polar}) can be well approximated by the isotropic
potential $\tilde{V}(r) \equiv \tilde{V}_0(r)$ given by Eq.~(\ref{V-isotropic}).

The lowest-order anisotropic correction to the potential
in Eq.~(\ref{V_expansion}) is given in terms of the Fourier coefficients,
\begin{eqnarray}
  \tilde{V}_{1}(r)
  =
  \tilde{V}_{-1}(r)
  \approx
  \frac{V_0}{180}~
  \bigg(\frac{\pi r}{\lambda_0}\bigg)^{6}.
  \label{V-anisotropic}
\end{eqnarray}
For $r=r_0=0.068 \, \lambda_0$, where $r_0$ is radius where the areal 
probability density is maximal,
$|\tilde{V}_1(r_0)| \approx 5.281 \times 10^{-7} \, V_0$,
i.e., it is really small, and therefore the anisotropic corrections can
be neglected for the ground and lowest energy eigenstates.

The isotropic potential $\tilde{V}_{0}(r)$ is shown in
Fig.~\ref{Fig-V_B_vs_r} (blue curve). Clearly,
$\tilde{V}_{0}(r)$ is attractive and increases
monotonically with $r$.

\subsection{Isotropic approximation for $\mbfB(\mbfr)$}

\begin{figure}
\centering
  \subfigure[]
   {\includegraphics[width = 0.9 \linewidth,angle=0]
   {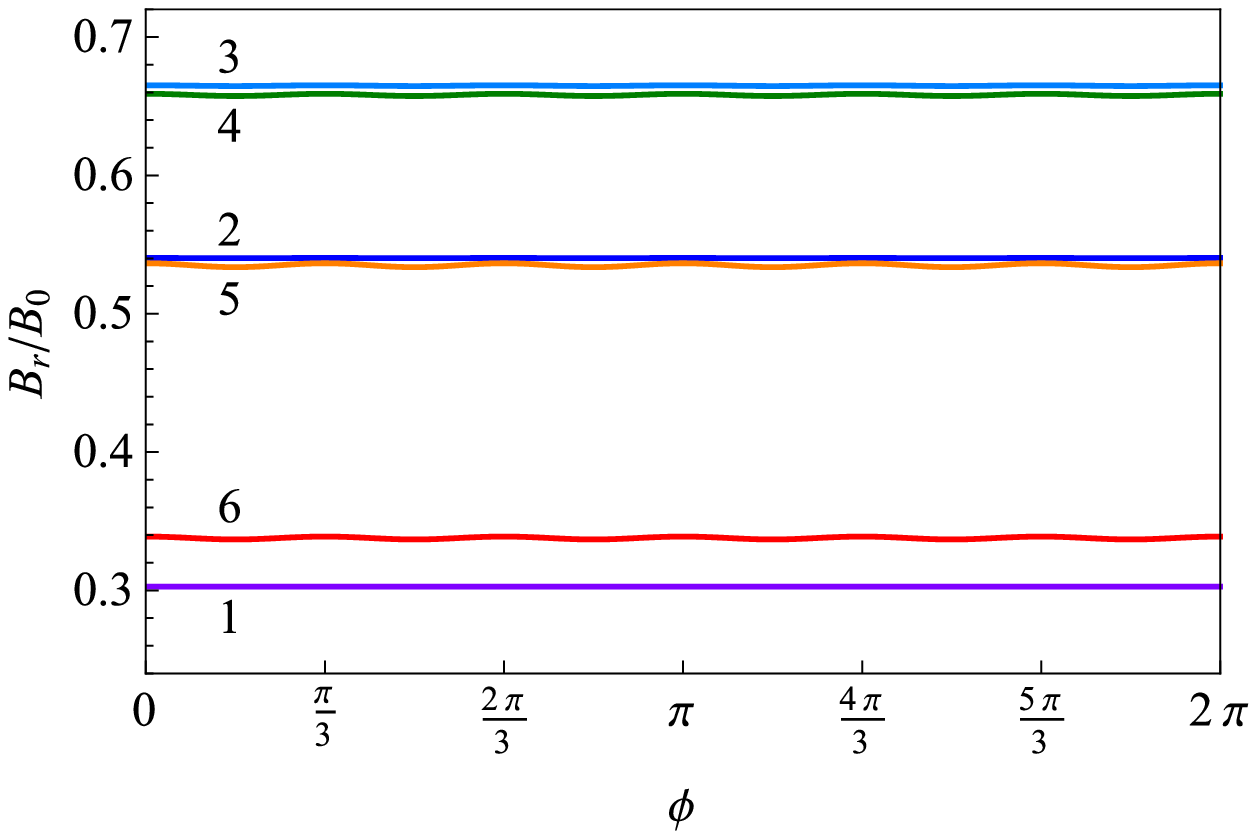}
   \label{Fig-Br-isotropic}}
  \subfigure[]
   {\includegraphics[width = 0.9 \linewidth,angle=0]
   {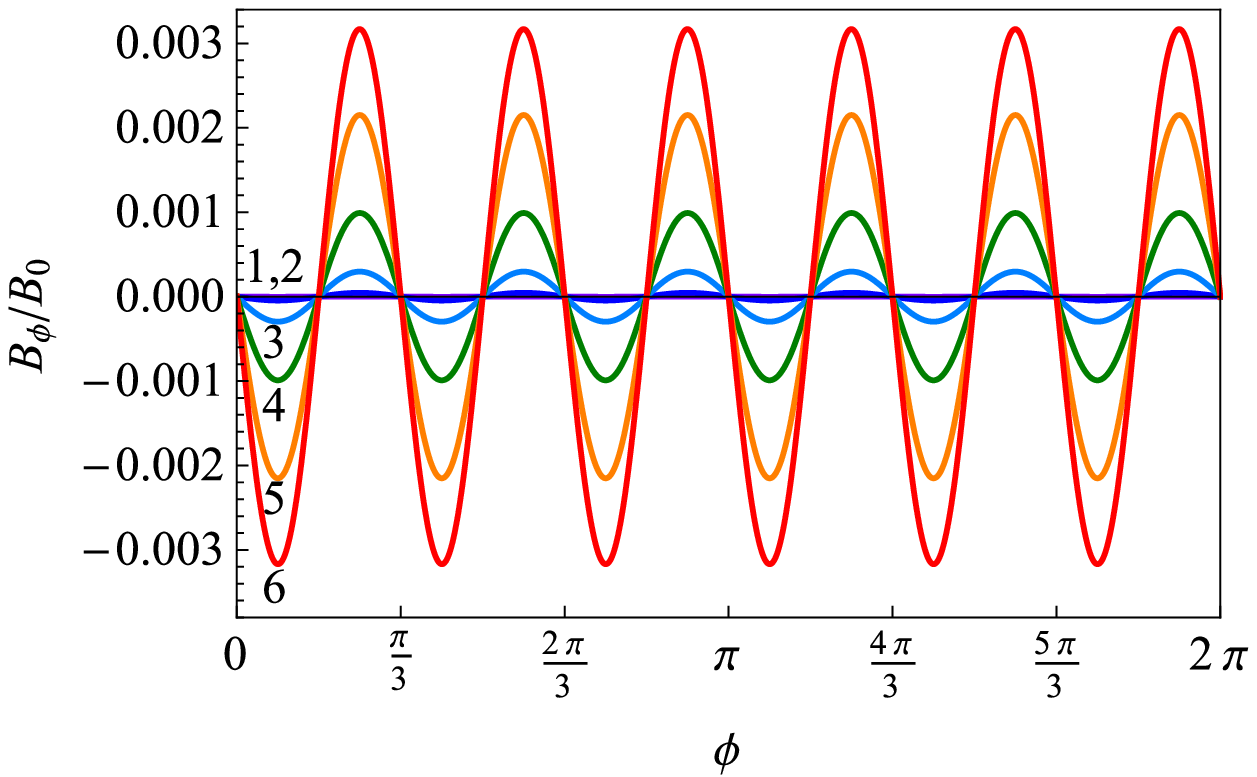}
   \label{Fig-Bp-isotropic}}
 \caption{
   (a) $B_r(\mbfr)$ and (b) $B_{\phi}(\mbfr)$,
   Eq.~(\ref{subeqs-Br-Bphi-polar}),
   as functions of the polar angle
   $\phi$ for $\beta=\frac{1}{\sqrt{2}}$.
   The purple (1), blue (2), sky blue (3),
   green (4), orange (5) and red (6)
   curves denote $r=0.05 \lambda_0$,
   $0.1 \lambda_0$, $0.15 \lambda_0$,
   $0.2 \lambda_0$, $0.25 \lambda_0$
   and $0.3 \lambda_0$.}
 \label{Fig-Br-Bp-isotropic}
\end{figure}

\begin{figure}
\centering
   {\includegraphics[width=\linewidth,angle=0]
   {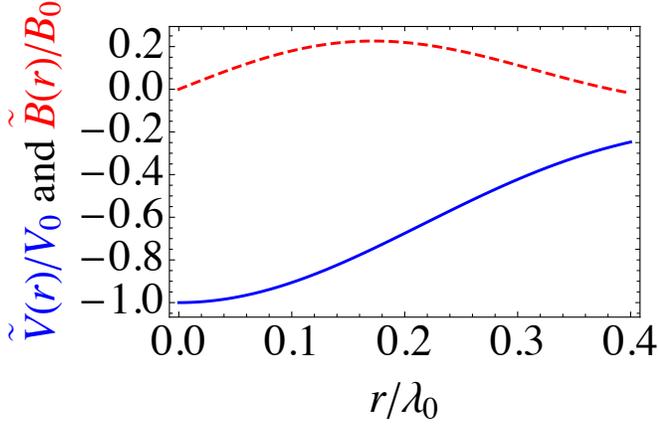}}
  \caption{
  Spherically symmetric 
  potential $\tilde{V}(r) \equiv \tilde{V}_0(r)$ in units of $V_0$ and the effective 
  radial magnetic field $\tilde{B}(r) \equiv \tilde{B}_{r,0}(r)$ in units of $B_0$.}
  \label{Fig-V_B_vs_r}
\end{figure}

Figure~\ref{Fig-Br-Bp-isotropic} shows $B_r(\mbfr)$ and
$B_{\phi}(\mbfr)$, Eq.~(\ref{subeqs-Br-Bphi-polar}), as functions
of $\phi$ for $\beta=1/\sqrt{2}$ and several values of $r$.
Clearly, for $r<0.3 \lambda_0$, $B_r(\mbfr)$ depends on on 
$r$, whereas $B_{\phi}(\mbfr)$ oscillates quickly with $\phi$, but 
with very small amplitude. Hence, for $r \lesssim 0.3 \lambda_0$, 
$B_r(\mbfr)$ can be approximated by the isotropic function
$\tilde{B}(r) \equiv \tilde{B}_{r,0}(r)$
\begin{eqnarray}
  \tilde{B}_{r,0}(r) &=&
  \frac{B_0}{3 (2I+1)}
  \Bigg\{
       J_1
       \bigg(
            \frac{2 \pi r}{\lambda_0}
       \bigg)+
       J_1
       \bigg(
            \frac{4 \pi r}{\lambda_0}
       \bigg)+
  \nonumber \\ && +
       \sqrt{3}~
       J_1
       \bigg(
            \frac{2 \sqrt{3} \pi r}{\lambda_0}
       \bigg)
  \Bigg\},
  \label{Br-isotropic}
\end{eqnarray}
and the isotropic part of $\tilde{B}_{\phi,0}(r)$
vanishes,
$$
  \tilde{B}_{\phi,0}(r) = 0.
$$
The lowest-order anisotropic corrections to $B_r(\mbfr)$ and
$B_{\phi}(\mbfr)$ in Eqs.~(\ref{Fourier-B-r})-(\ref{Fourier-B-phi}) 
are given by the Fourier coefficients
\begin{eqnarray}
  \tilde{B}_{r,1}(r) = \tilde{B}_{r,-1}(r)
  \approx
  -\frac{B_0}{120 (2I+1)}~
  \bigg(\frac{\pi r}{\lambda_0}\bigg)^{5},
  \label{B-r-anisotropic}
  \\
  \tilde{B}_{\phi,1}(r)
  =
  \tilde{B}_{\phi,-1}^{*}(r)
  \approx
  -\frac{i B_0}{120 (2I+1)}~
  \bigg(\frac{\pi r}{\lambda_0}\bigg)^{5}.
  \label{B-ph-anisotropic}
\end{eqnarray}
When $r = r_0 = 0.068 \, \lambda_0$,
$\big| \tilde{B}_{r,1} \big(r_0\big) \big| =
\big| \tilde{B}_{\phi,1} \big(r_0\big) \big|
\approx 1.236 \times 10^{-6}~B_0$,
is very small, and the anisotropic part of the magnetic
field can be neglected for the ground and lowest excited 
energy eigenstates.

The isotropic effective radial magnetic field $\tilde{B}(r)$
is shown in Fig.~\ref{Fig-V_B_vs_r} (red curve). Note
that $\tilde{B}(0)=0$.
There is a distance $r_B=0.1722 \, \lambda_0$
such that for $r<r_B$, $\tilde{B}(r)$ increases
with increasing $r$.
$\tilde{B}(r)$ reaches its maximum,
$\tilde{B}(r_B)=0.6780 \, B_0$, at $r=r_B$,
and decreases for $r>r_B$. When
$r=r_c=0.3827 \, \lambda_0$,
$\tilde{B}(r_c)=0$, and for $r>r_c$,
the fictitious magnetic field reverses its direction
from $\hat{\mbfr}$ to $-\hat{\mbfr}$.

\section{Probability density for $F = 1/2$}  \label{Sec:probability}

At finite temperature $T$, the probability density to find the atom
at position ${\bf r}$ in the 2D plane is
$$
  \rho(r) =
  \frac{1}{Z}
  \sum_{n,\zeta}
  \big|\Psi_{n,\zeta}(\mbfr)\big|^2~
  e^{-\beta \epsilon_{n,\zeta}},
$$
where
$$
  Z =
  \sum_{n,\zeta}
  e^{-\beta \epsilon_{n,\zeta}},
$$
and $\beta = (k_B T)^{-1}$ is proportional to the inverse temperature.
At low temperature, where $k_B T$ is much smaller 
than the orbital excitation energy [see Eq.~(\ref{energy-orbit})],
$\rho(r)$ is well approximated by Eq.~(\ref{density-low-T}).
In the isotropic optical lattice potential approximation, $\rho(r)$
depends only on $r$ and not on $\phi$.  $\rho(r)$ has a maximum at 
$r=r_0=0.068 \, \lambda_0$, and rapidly decays with $|r-r_0|$, as shown in 
Fig.~\ref{Fig-density}.

\section{Semiclassical Description of the Quantum Rotor}
  \label{Sec:Semiclassical}

When $F$ is large, $F \gg 1$,
we can describe the motion of the trapped atoms using
a semiclassical approximation.  This formulation yields useful insights
into the behavior of QRs, and is simple to carry out
within the isotropic approximation for the SDOLP, Eqs.~(\ref{V-isotropic}) 
and (\ref{Br-isotropic}), 
which is valid near the minimum of the scalar potential.  A power series
expansion about the origin yields:
\begin{eqnarray}
&&  \tilde{V}_0(r) \approx
  -V_0+
  \frac{\pi^2 V_0 r^2}{\lambda_0^2}, 
  \nonumber  \\
&&  \tilde{B}_{r,0}(r) \approx
  \frac{2 \pi B_0 r}{(2 I + 1)\lambda_0}.
  \label{Power-series}
\end{eqnarray}
In terms of the canonical momentum $\mbfp$ and the unit vector 
${\mathbf{f}}=\mbfF/F$, and using the power series in Eqs.~(\ref{Power-series}),
we arrive at the following Hamilton equations of motion,
\begin{subequations}
   \label{subeqs-r-p-F-Hamilton}
\begin{eqnarray}
  \dot\mbfr &=&
  \frac{\mbfp}{M},
  \label{eq1-r-Hamilton}
  \\
  \dot\mbfp &=&
  -\frac{2 \pi^2 V_0}{\lambda_0^2}~
  \mbfr-
  \frac{2 \pi F B_0}{(2 I + 1) \lambda_0}~
  {\mathbf{f}}_{\parallel},
  \label{eq2-p-Hamilton}
  \\
  \dot{\mathbf{f}} &=&
  \frac{2 \pi B_0}{(2 I + 1) \hbar \lambda_0}~
  \big[
      \mbfr
      \times
      {\mathbf{f}}
  \big],
  \label{eq3-F-Hamilton}
\end{eqnarray}
\end{subequations}
where ${\mathbf{f}}_{\parallel} = f_x \mbfe_x + f_y \mbfe_y$.

\begin{figure}
\centering
  \subfigure[]
  {\includegraphics[width=0.9\linewidth,angle=0]
   {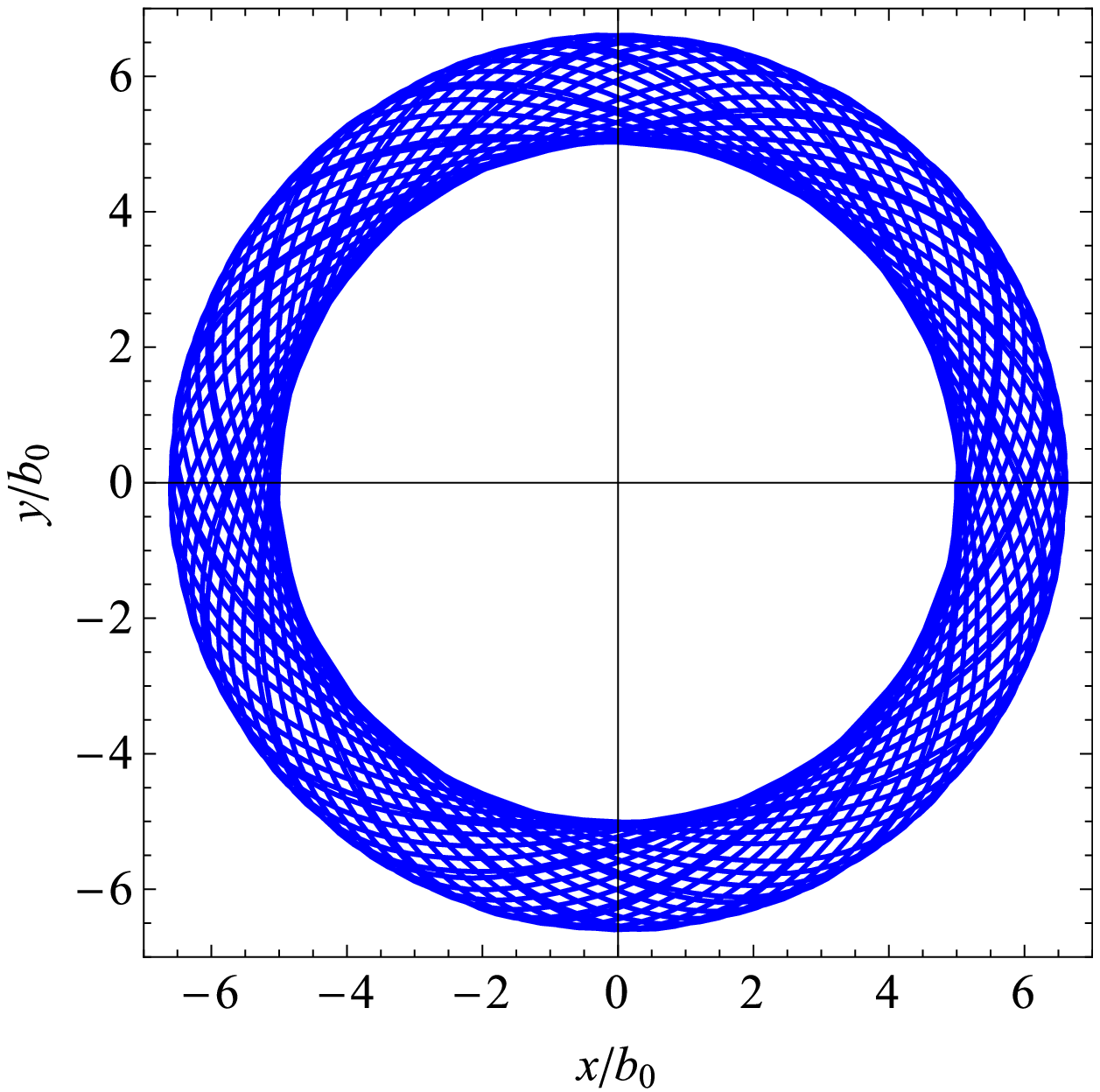}
   \label{Fig-XY}}
  \subfigure[]
  {\includegraphics[width=0.9\linewidth,angle=0]
   {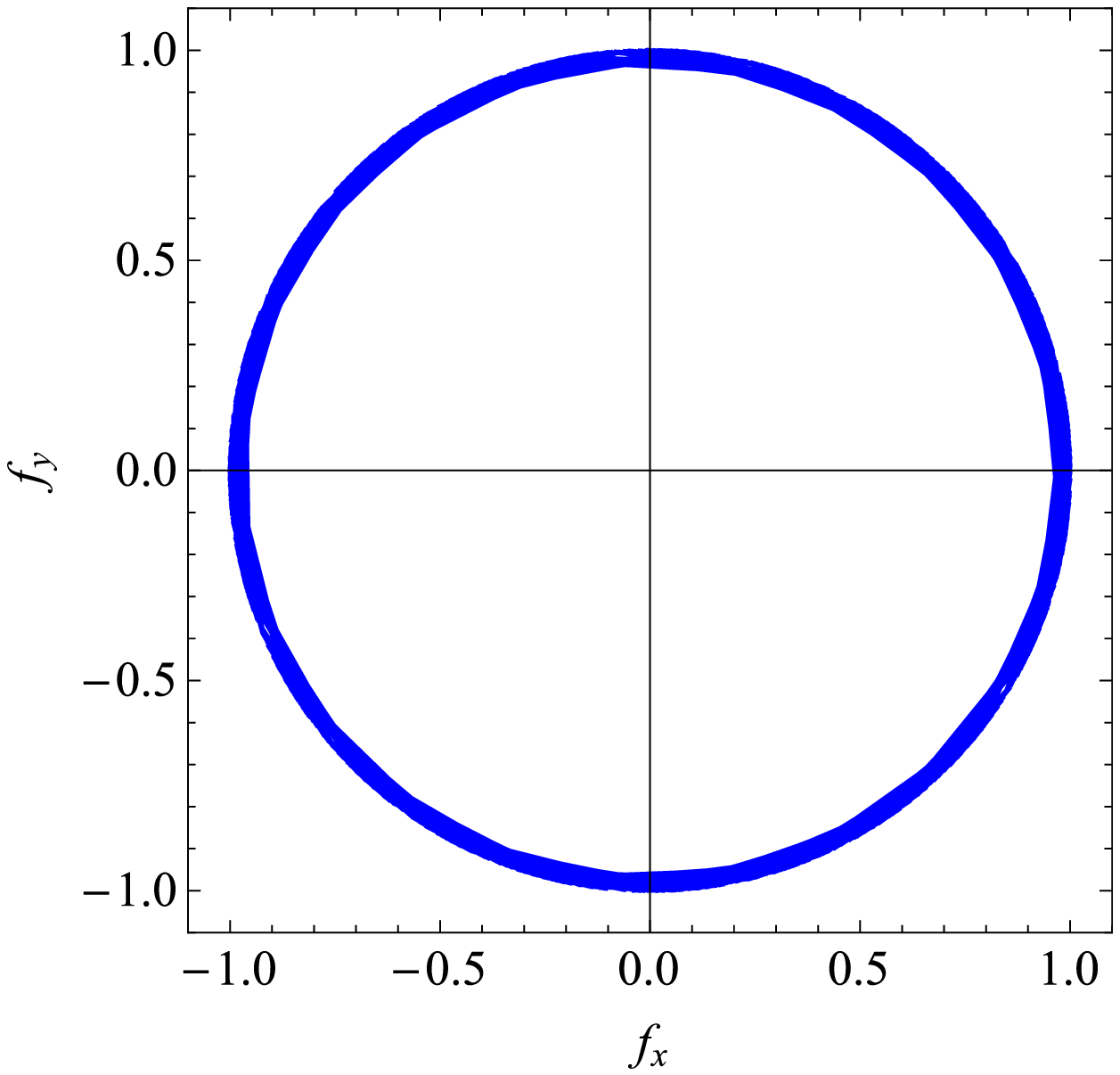}
   \label{Fig-FxFy}}
 \caption{
   (a) Position of the trapped atom ${\bf r}(t) = (x(t),y(t))$ and 
   (b) the unit vector ${\bf f}(t) = (f_x(t),f_y(t))$ parallel to
   the atomic orbital momentum.  Here, we have taken
   $B_0=1.8 V_0$ and initial conditions (initial position,
   momentum and the direction of the total atomic angular
   momentum), ${\bf f}(t) = {\bf F}(t)/|{\bf f}(t)|$ given by
   Eq.~(\ref{initial-num}).
   }
 \label{Fig-XY-FxFy}
\end{figure}

It is convenient to use harmonic length and energy units:
\begin{eqnarray*}
  &&
  {\text{harmonic length}}
  \ \ \ \ \
  \ \ \ \
  b_0
  =
  \frac{\lambda_0}{2 \pi}~
  \bigg(
       \frac{2 {\mathcal{E}}_{0}}{V_0}
  \bigg)^{1/4},
  \\
  &&
  {\text{harmonic energy}}
  \ \ \ \ \
  \hbar \Omega_h =
  \sqrt{\frac{{\mathcal{E}}_{0} V_0}{2}},
\end{eqnarray*}
where the recoil energy ${\mathcal{E}}_{0}$ is
\begin{eqnarray}
  {\mathcal{E}}_{0} =
  \frac{\hbar^2 q_{0}^{2}}{2 M} =
  \frac{ 2 \pi^2 \hbar^2}{M \lambda_{0}^{2}}.
  \label{recoil-energy-def}
\end{eqnarray}

In this section, we consider $^{40}$K atoms which have
$I=4$, $J=1/2$ and $F=9/2$ in the ground state.
When $V_0 = 100 \, {\mathcal{E}}_{0}$, 
$b_0=0.05985~\lambda_0$ and $\hbar\Omega_h = 7.071 \, {\mathcal{E}}_{0}$.
The time-dependent position of the atom ${\bf r}(t) = (x(t),y(t))$ and
the unit spin vector ${\bf f}(t) = (f_x(t),f_y(t))$ are shown in
Fig.~\ref{Fig-XY-FxFy} for $B_0=1.8 \, V_0$, with the following initial conditions,
\begin{eqnarray}
  &&
  x(0) = 5~b_0,
  \ \ \ \ \
  y(0) = 0,
  \nonumber
  \\
  &&
  p_x(0) = 0,
  \ \ \ \ \
  p_y(0) = -3 \hbar/b_0,
  \label{initial-num}
  \\
  &&
  f_x(0) = -0.9975,
  \ \ \
  f_y(0) = 0.05,
  \ \ \ \ \
  f_z(0) = 0.05.
  \nonumber
\end{eqnarray}
The motion of the atom is not periodic, and therefore the trajectory 
$(x(t),y(t))$ in Fig.~\ref{Fig-XY} fills a ring with inner
radius $5~b_0$ and outer radius $6.615~b_0$, 
whereas $(f_x(t),f_y(t))$ in Fig.~\ref{Fig-FxFy} fills 
a ring with inner radius $0.9644$ and outer radius $1$.
For $^{40}$K atoms, $\lambda_0 = 766.5$~nm, and
$b_0 = 45.87$~nm.

The results of this section cannot be applied to Li atoms because,
for the semiclassical approximation to be valid, $F$ must be large.

\section{Distinguishing Between ${\boldsymbol \Omega}$, ${\bf a}$, ${\bf B}_{\mathrm{ex}}$} 
\label{Sec:Distinguishing}

Consider a QR placed in an external magnetic field ${\bf B}_{\mathrm{ex}}$
in a non-inertial frame moving with a linear acceleration $\mbfa$
and rotating with an angular velocity $\boldsymbol\Omega$.
The 3D Hamiltonian of the QR is
\begin{eqnarray}
  H &=& H_0 + H_z + H_B + H_{\Omega} + H_a.
  \label{H=H0+HB+Hrot+Ha}
\end{eqnarray}
Here $H_0$ is a 2D Hamiltonian for motion of the QR in the $x$-$y$ plane
without an external magnetic field and in an inertial frame; using polar coordinates 
$\mbfr=(r,\phi)$, $H_0 = -\frac{\hbar^2 \nabla^2}{2 M} + {\tilde V}(r) - \tilde{B}(r)F_r$.
The second term on the right hand side of Eq.~(\ref{H=H0+HB+Hrot+Ha}), $H_z$, is for the 
motion of the trapped atom in a harmonic potential in the $z$-direction,
\begin{eqnarray}
  H_z &=&
  -\frac{\hbar^2 \partial_{z}^{2}}{2 M} + \frac{K_z z^2}{2}.
  \label{Hz-def}
\end{eqnarray}
The harmonic oscillator force constant $K_z$ is assumed to be large and 
the atom is in the ground state of the Hamiltonian (\ref{Hz-def}).  The third term 
on the right hand side of Eq.~(\ref{H=H0+HB+Hrot+Ha}), $H_B$, is the Zeeman 
interaction between the QR and the external magnetic field ${\bf B}_{\mathrm{ex}}$, 
$H_B = -\frac{g \mu_B}{2 I + 1}~\mbfF \cdot {\bf B}_{\mathrm{ex}}$.  The fourth term, 
$H_{\Omega}$, is due to the fictitious force in a rotating frame of reference,
$H_{\Omega} = \hbar \boldsymbol{\mathcal{L}} \cdot \boldsymbol\Omega$,
where $\boldsymbol{\mathcal{L}}=\mbfF+\boldsymbol{\ell}$, and
$\boldsymbol{\ell}$ is the orbital angular momentum of the atom around
minimum points of $\tilde{V}(r)$.  The fifth term, $H_a$, is due to the 
fictitious force appearing in a 
non-inertial frame moving with linear acceleration $\mbfa$, $H_a = M \mbfa \cdot \mbfr$.

We shall calculate energy levels of $H_0 + H_z$ in the inertial frame, and
then apply first order perturbation theory in ${\bf B}_{\mathrm{ex}}$,
$\boldsymbol\Omega$ and $\mbfa$, to find the corrections to the energies of the QR.

\subsection{Matrix elements of $H_B$}

Matrix elements of $F_x$, $F_y$ and $F_z$ are
\begin{eqnarray*}
  \big\langle
      n,\zeta
  \big|
      F_x
  \big|
      n,\zeta'
  \big\rangle
  &=&
  \frac{1}{2}~
  \beta_{n,\zeta}^{\parallel}~
  \big(
      \delta_{\zeta,\zeta'+1}+
      \delta_{\zeta,\zeta'-1}
  \big),
  \\
  \big\langle
      n,\zeta
  \big|
      F_y
  \big|
      n,\zeta'
  \big\rangle
  &=&
  -\frac{i}{2}~
  \beta_{n,\zeta}^{\parallel}~
  \big(
      \delta_{\zeta,\zeta'+1}-
      \delta_{\zeta,\zeta'-1}
  \big),
  \\
  \big\langle
      n,\zeta
  \big|
      F_z
  \big|
      n,\zeta'
  \big\rangle
  &=&
  \beta_{n,\zeta}^{z}
  \delta_{\zeta,\zeta'},
\end{eqnarray*}
where $\langle z, \mbfr | n,\zeta\rangle = \tilde\psi_{0}(z) \Psi_{n,\zeta}(\mbfr)$,
and
\begin{eqnarray}
  \beta_{n,\zeta}^{\parallel} &=&
  \frac{1}{2}
  \int
  \Big\{
      \psi_{n,\zeta,1/2}^{2}(r) -
      \psi_{n,\zeta,-1/2}^{2}(r)
  \Big\}~
  dr,
  \label{beta-xy-def}
  \\
  \beta_{n,\zeta}^{z} &=&
  \int
  \psi_{n,\zeta,1/2}(r)
  \psi_{n,\zeta,-1/2}(r)~
  dr.
  \label{beta-z-def}
\end{eqnarray}
Note the following symmetries:
$$
  \beta_{n,-\zeta}^{\parallel}
  =
  \beta_{n,\zeta}^{\parallel},
  \qquad
  \beta_{n,-\zeta}^{z}
  =
  -\beta_{n,\zeta}^{z}.
$$
These follow from the wave function symmetries
$\psi_{n,-\zeta,1/2}(r) = \psi_{n,\zeta,1/2}(r)$
and
$\psi_{n,-\zeta,-1/2}(r) = -\psi_{n,\zeta,-1/2}(r)$.
The matrix elements of $H_B$ are
\begin{eqnarray}
  &&
  \big\langle
      n,\zeta
  \big|
      H_B
  \big|
      n,\zeta'
  \big\rangle
  =
  -\frac{g \mu_B}{2 I+1}~
  \bigg\{
       \zeta B_z \beta_{n,\zeta}^{z} \delta_{\zeta,\zeta'} +
  \nonumber \\ && ~~~~~ ~~~~~ ~~~~~
       \frac{\beta_{n,\zeta}^{\parallel}}{2}~
       \Big[
           B^{+}
           \delta_{\zeta,\zeta'-1} +
           B^{-}
           \delta_{\zeta,\zeta'+1}
       \Big]
  \bigg\},
  \label{HB-ME}
\end{eqnarray}
where $B^{\pm} = B_x \pm i B_y$.

\subsection{Matrix elements of $H_{\Omega}$}

Matrix elements of the operator $\boldsymbol{\mathcal{L}}$ are
\begin{eqnarray*}
  &&
  \big\langle
      n,\zeta
  \big|
      {\mathcal{L}}_{x}
  \big|
      n,\zeta'
  \big\rangle
  =
  \big\langle
      n,\zeta
  \big|
      {\mathcal{L}}_{y}
  \big|
      n,\zeta'
  \big\rangle
  = 0,
  \\
  &&
  \big\langle
      n,\zeta
  \big|
      {\mathcal{L}}_{z}
  \big|
      n,\zeta'
  \big\rangle
  =
  \zeta \Omega_z.
\end{eqnarray*}
The matrix elements of ${\mathcal{L}}_{x}$ and ${\mathcal{L}}_{y}$
vanish since the wave function $\tilde\psi_{0}(z)$ is even with
respect to the inversion $z \to -z$, whereas ${\mathcal{L}}_{x}$
and ${\mathcal{L}}_{y}$ are odd. Hence, non-vanishing matrix 
elements of $H_{\Omega}$ are
\begin{eqnarray}
  \big\langle
      n,\zeta
  \big|
      H_{\Omega}
  \big|
      n,\zeta'
  \big\rangle
  &=&
  \hbar \zeta \Omega_z \delta_{\zeta,\zeta'}.
  \label{HOmega-ME}
\end{eqnarray}

\subsection{Matrix elements of $H_a$}

Matrix elements of the position operator $\mbfr$ are
\begin{eqnarray*}
  \big\langle
      n,\zeta
  \big|
      x
  \big|
      n,\zeta'
  \big\rangle
  &=&
  \frac{1}{2}~
  \varrho_{n,\zeta}^{\parallel}~
  \big(
      \delta_{\zeta,\zeta'+1}+
      \delta_{\zeta,\zeta'-1}
  \big),
  \\
  \big\langle
      n,\zeta
  \big|
      y
  \big|
      n,\zeta'
  \big\rangle
  &=&
  -\frac{i}{2}~
  \varrho_{n,\zeta}^{\parallel}~
  \big(
      \delta_{\zeta,\zeta'+1}-
      \delta_{\zeta,\zeta'-1}
  \big),
  \\
  \big\langle
      n,\zeta
  \big|
      z
  \big|
      n,\zeta'
  \big\rangle
  &=& 0,
\end{eqnarray*}
where
\begin{eqnarray}
  \varrho_{n,\zeta}^{\parallel} &=&
  \int
  \Big\{
      \psi_{n,\zeta,1/2}^{2}(r) +
      \psi_{n,\zeta,-1/2}^{2}(r)
  \Big\}~
  r dr.
  \label{alpha-xy-def}
\end{eqnarray}
Note the symmetry $\varrho_{n,-\zeta}^{\parallel} =
\varrho_{n,\zeta}^{\parallel}$.
The matrix elements of $H_a$ are
\begin{eqnarray}
  \big\langle
      n,\zeta
  \big|
      H_a
  \big|
      n,\zeta'
  \big\rangle
  =
  \frac{M \varrho_{n,\zeta}^{\parallel}}{2}~
  \Big[
      a^{+}
      \delta_{\zeta,\zeta'-1} +
      a^{-}
      \delta_{\zeta,\zeta'+1}
  \Big],
  \label{Ha-ME}
\end{eqnarray}
where $a^{\pm} = a_x \pm i a_y$.

\subsection{First-order corrections to the energies}  \label{SubSec:FOCE}

In order to find the first order corrections to the energies of the QR
due to $\mbfB$, $\boldsymbol\Omega$ and $\mbfa$,
we apply degenerate perturbation theory.
First order corrections to the energies of the quantum
states with $\zeta = \pm 1/2$ are
\begin{eqnarray}
  &&
  \veps_{n,\pm 1/2}^{(1)}(\mbfB,\Omega_z,a_{\parallel})
  =
  \pm \frac{1}{2}
  \Bigg[
       \bigg(
            \frac{g \mu_B \beta_{n,1/2}^{z}}{2 I+1}~B_z +
            \hbar \Omega_z
       \bigg)^{2} +
  \nonumber \\ && ~~~~~ ~~~~~ ~~~~~
       \Big|
           \beta_{n,1/2}^{\parallel} B^{+} +
           M \varrho_{n,1/2}^{\parallel} a^{+}
       \Big|^{2}
  \Bigg]^{1/2}.
  \label{energy-pm-1/2-B-Omega-a}
\end{eqnarray}
Corrections to the energies of the quantum states with
$\pm\zeta$ (where $\zeta = 3/2, 5/2, 7/2, \ldots$) are
\begin{eqnarray}
  \veps_{n,\pm \zeta}^{(1)}(\mbfB,\Omega_z,a_{\parallel})
  &=&
  \pm \zeta
  \bigg(
       \frac{g \mu_B \beta_{n,\zeta}^{z}}{2 I+1}~B_z +
       \hbar \Omega_z
  \bigg).
  \label{energy-pm-zeta-B-Omega-a}
\end{eqnarray}

\subsection{Raman Spectroscopy Considerations for Distinguishing between
Various Sensors} \label{SubSec:Raman_Sensors}

We propose to use Raman spectroscopy to measure $\mbfB$, 
$\boldsymbol\Omega$ and $\mbfa$ by applying
radio frequency electromagnetic waves to the QR with pump and
Stokes frequencies ($\omega_p > \omega_s$) that are far-off-resonance
from the $F=3/2$ atomic hyperfine state.  In Sec.~\ref{SubSec:Raman}
we discussed the Raman transition $|0,1/2\rangle \leftrightarrow |0,-1/2\rangle$ 
and the use of the Ramsey separated oscillating fields method to
verify the Raman resonance condition $\omega \equiv \omega_p - \omega_s
= \Delta_{QR}$.  In order to determine ${\boldsymbol \Omega}$, 
${\bf a}$, and ${\bf B}_{\mathrm{ex}}$, we will need to consider the 
Raman transitions $|0,1/2\rangle \leftrightarrow |0,-1/2\rangle$,
$|0,3/2\rangle \leftrightarrow |0,5/2\rangle$ and $|0,-3/2\rangle 
\leftrightarrow |0,-5/2\rangle$, which have transition frequencies $\Delta_0 
\equiv \Delta_{QR}$, $\Delta_{+}$ and $\Delta_{-}$ respectively:
\begin{eqnarray*}
  \Delta_0 &=&
  \frac{1}{\hbar}~
  \Big(
      \veps_{0,1/2}^{(1)}
      \big(\mbfB,\Omega_z,a_{\parallel}\big) -
      \veps_{0,-1/2}^{(1)}
      \big(\mbfB,\Omega_z,a_{\parallel}\big)
  \Big),
  \\
  \Delta_{\pm} &=&
  \frac{1}{\hbar}
  \Big(
      \epsilon_{0,5/2} -
      \epsilon_{0,3/2} +
      \veps_{0,\pm 5/2}^{(1)}
      \big(\mbfB,\Omega_z,a_{\parallel}\big)
  \nonumber \\ && -
      \veps_{0,\pm 3/2}^{(1)}
      \big(\mbfB,\Omega_z,a_{\parallel}\big)
  \Big).
\end{eqnarray*}
Note that when $\mbfB_{\mathrm{ex}}=0$, $\boldsymbol\Omega=0$
and $\mbfa=0$, then
$$
  \Delta_{+}^{(0)} =
  \Delta_{-}^{(0)} =
  \frac{1}{\hbar}~
  \big(
      \epsilon_{0,5/2} -
      \epsilon_{0,3/2}
  \big).
$$
We are interested in the splitting $\Delta_1 = \Delta_{+} - \Delta_{-}$
due to $\mbfB_{\mathrm{ex}}$, $\boldsymbol\Omega$ and $\mbfa$.
Using Eqs.~(\ref{energy-pm-1/2-B-Omega-a}) and
(\ref{energy-pm-zeta-B-Omega-a}), we get
\begin{eqnarray}
  &&
  \Delta_0
  \Big(
      B_x,B_y,B_z,\Omega_z,a_x,a_y;
      \varrho_{0,1/2}^{\parallel},
      \beta_{0,1/2}^{\parallel},
      \beta_{0,1/2}^{z}
  \Big)
  =
  \nonumber \\ && ~~~~~ ~~~~~
  \frac{1}{\hbar}~
  \Bigg[
       \bigg(
            \frac{g \mu_B \beta_{0,1/2}^{z}}{2 I+1}~B_z +
            \hbar \Omega_z
       \bigg)^{2}  +
  \nonumber \\ && ~~~~~ ~~~~~ ~~
       \bigg(
            \frac{g \mu_B \beta_{0,1/2}^{\parallel}}{2 I + 1}~B_x +
            M \varrho_{0,1/2}^{\parallel} a_x
       \bigg)^{2}
  \nonumber  +\\ && ~~~~~ ~~~~~ ~~
       \bigg(
            \frac{g \mu_B \beta_{0,1/2}^{\parallel}}{2 I + 1}~B_y +
            M \varrho_{0,1/2}^{\parallel} a_y
       \bigg)^{2}
  \Bigg]^{1/2},
  \label{Omega0}
  \\
  &&
  \Delta_1
  \Big(
      B_z,\Omega_z;
      \beta_{0,3/2}^{z},\beta_{0,5/2}^{z}
  \Big)
  =
  \frac{1}{\hbar}~
  \bigg\{
       2 \hbar \Omega_z +
  \nonumber \\ && ~~~~~ ~~~~~ ~~
       \frac{g \mu_B}{2 I+1}
       \Big(
           5 \beta_{0,5/2}^{z} -
           3 \beta_{0,3/2}^{z}
       \Big)
       B_z
  \bigg\}.
\end{eqnarray}

Nine measurements need to be made to allow determination of the 9 unknowns: 
$B_z$, $B_y$, $B_z$, $\Omega_x$, $\Omega_y$, $\Omega_z$,
$a_x$, $a_y$ and $a_z$. In order to find the 9 unknowns,
9 measurements are required. In particular, measurements must be
carried out with the QR placed in $x$-$y$,
$y$-$z$ and $z$-$x$ optical lattices.
Moreover, measurements of $\Delta_1$
must be made with two different laser intensities, e.g.,
$(V_0,B_0) = (100 \, {\mathcal{E}}_{0}, 180 \, {\mathcal{E}}_{0})$
and $(50 \, {\mathcal{E}}_{0},90 \, {\mathcal{E}}_{0})$.
In other words, we consider $(V_0,B_0) = 
(10 \, {\mathcal{N}} {\mathcal{E}}_{0}, 18 \, {\mathcal{N}} {\mathcal{E}}_{0})$,
where the dimensionless parameter ${\mathcal{N}} = 5, 10$
specifies the laser intensity.  Furthermore, measurements of $\Delta_0$
must be made with $V_0=100 \, {\mathcal{E}}_{0}$
and $B_0 = 180 \, {\mathcal{E}}_{0}$.

\begin{figure}
\centering
  \subfigure[]
   {\includegraphics[width=0.60\linewidth,angle=0]
   {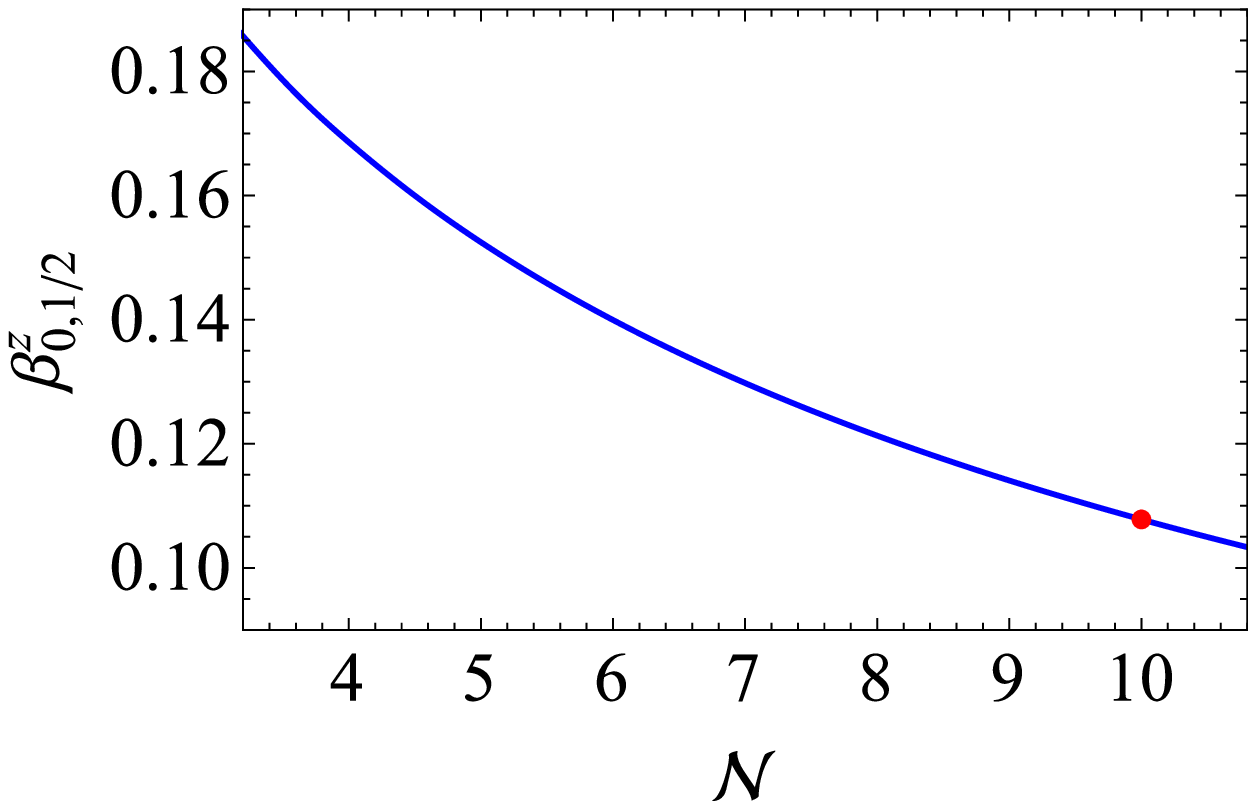}
   \label{Fig-betaZ1}}
  \subfigure[]
   {\includegraphics[width=0.60\linewidth,angle=0]
   {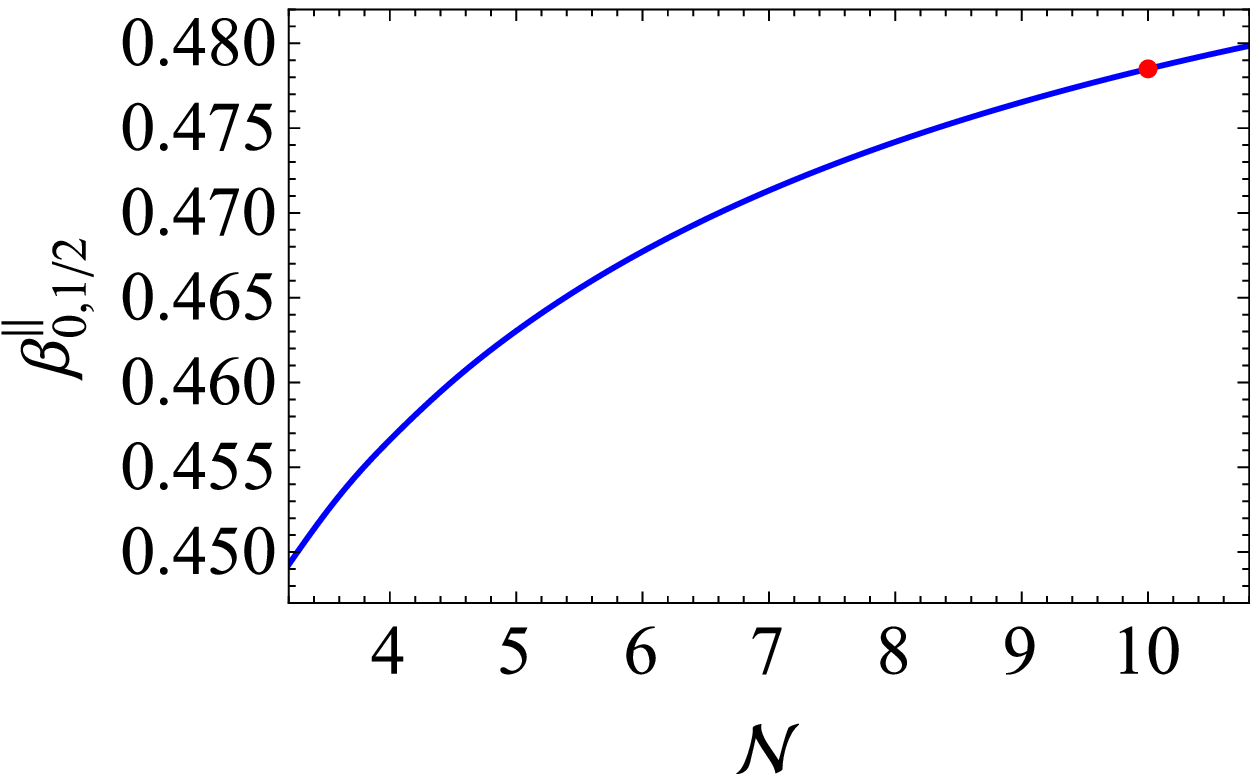}
   \label{Fig-betaXY1}}
  \subfigure[]
   {\includegraphics[width=0.60\linewidth,angle=0]
   {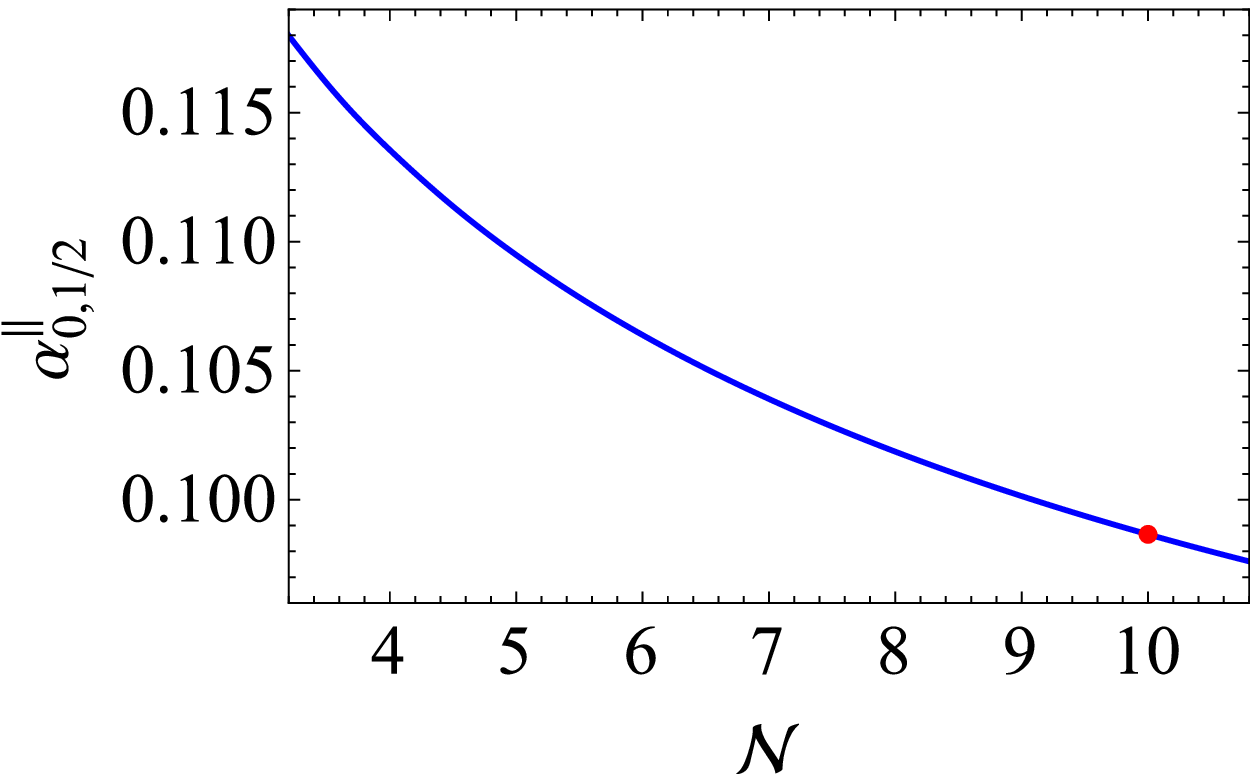}
   \label{Fig-alphaXY1}}
  \subfigure[]
   {\includegraphics[width=0.60\linewidth,angle=0]
   {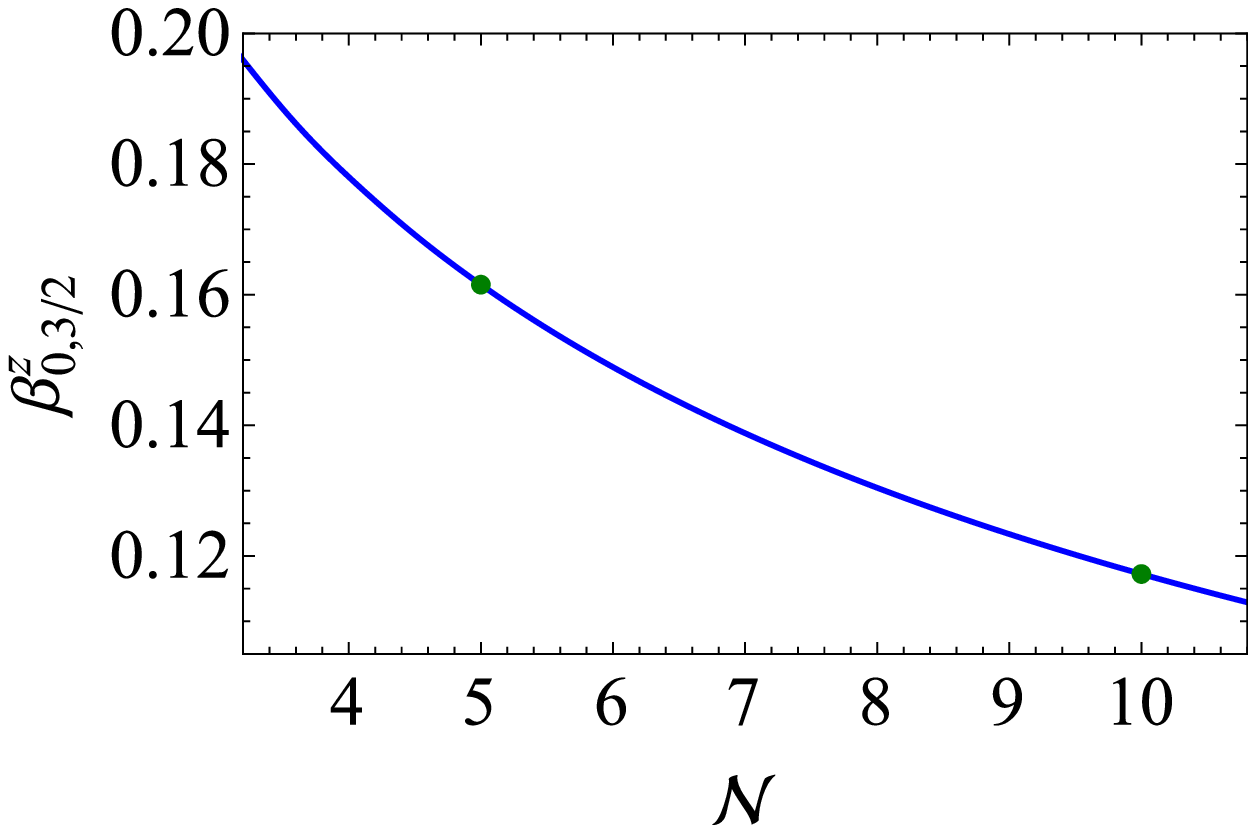}
   \label{Fig-betaZ3}}
  \subfigure[]
   {\includegraphics[width=0.60\linewidth,angle=0]
   {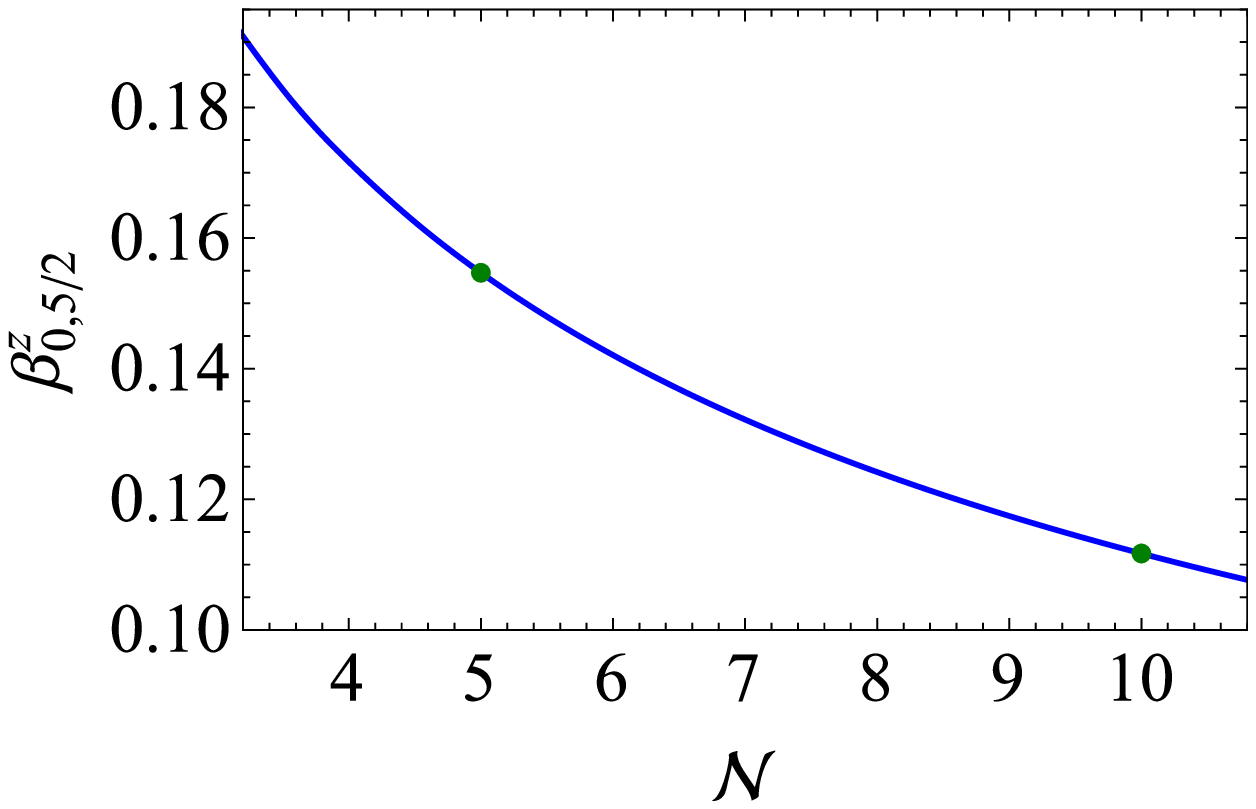}
   \label{Fig-betaZ5}}
 \caption{(a) $\beta_{0,1/2}^{z}$ (\ref{beta-z-def}),
 (b) $\beta_{0,1/2}^{\parallel}$ (\ref{beta-xy-def}),
 (c) $\varrho_{0,1/2}^{\parallel}$ (\ref{alpha-xy-def}),
 (d) $\beta_{0,3/2}^{z}$ (\ref{beta-z-def}) and
 (e) $\beta_{0,5/2}^{z}$ (\ref{beta-z-def})
 as functions of ${\mathcal{N}}$.
 The red dots denote the values of
 $\beta_{0,1/2}^{z}$, $\beta_{0,1/2}^{\parallel}$ and
 $\varrho_{0,1/2}^{\parallel}$ for ${\mathcal{N}} = 10$,
 and the green dots denote the values of
 $\beta_{0,3/2}^{z}$ and $\beta_{0,5/2}^{z}$
 for ${\mathcal{N}} = 5$ and $10$.}
 \label{Fig-betaZ-betaXY-alphaXY}
\end{figure}

The numerical solution of the Schr\"odinger equation (\ref{WF-vs-WF-f-m}), 
for different values of ${\mathcal{N}}$ yields the following
results for the integrals $\beta_{0,1/2}^{z}({\mathcal{N}})$,
$\beta_{0,3/2}^{z}({\mathcal{N}})$,
$\beta_{0,5/2}^{z}({\mathcal{N}})$,
$\beta_{0,1/2}^{\parallel}({\mathcal{N}})$ and
$\varrho_{0,1/2}^{\parallel}({\mathcal{N}})$.
The values of $\beta_{0,3/2}^{z}({\mathcal{N}})$ and
$\beta_{0,5/2}^{z}({\mathcal{N}})$ for ${\mathcal{N}}=5,10$
are:
\begin{eqnarray*}
  &&
  \beta_{0,3/2}^{z}(5) ~=~
  0.161531,
  \ \ \ \ \
  \beta_{0,3/2}^{z}(10) ~=~
  0.117236,
  \\
  &&
  \beta_{0,5/2}^{z}(5) ~=~
  0.154678,
  \ \ \ \ \
  \beta_{0,5/2}^{z}(10) ~=~
  0.111702.
\end{eqnarray*}
The values of $\beta_{0,1/2}^{z}(10)$, $\beta_{0,1/2}^{\parallel}(10)$
and $\varrho_{0,1/2}^{\parallel}(10)$ for ${\mathcal{N}}=10$
are:
\begin{eqnarray*}
  &&
  \beta_{0,1/2}^{z}(10)
  ~=~
  0.107807,
  \\
  &&
  \beta_{0,1/2}^{\parallel}(10)
  ~=~
  0.478494,
  \\
  &&
  \varrho_{0,1/2}^{\parallel}(10)
  ~=~
  0.0986575~
  \lambda_0.
\end{eqnarray*}

\subsection{Determining ${\bf B}_{\mathrm{ex}}$, $\boldsymbol\Omega$ and $\mbfa$}

With the optical lattice in the $x$-$y$ plane,
measurements can be made of $\Delta_{1,xy}({\mathcal{N}})$
for ${\mathcal{N}}=5,10$. Two equations (one for ${\mathcal{N}}=5$ and one for
${\mathcal{N}}=10$) are thereby obtained from
\begin{eqnarray}
  &&
  \frac{g \mu_B B_z}{\hbar(2 I+1)}
  \Big(
      5 \beta_{0,5/2}^{z}({\mathcal{N}}) -
      3 \beta_{0,3/2}^{z}({\mathcal{N}})
  \Big)
  \nonumber \\ && ~~~~~ ~~~~~ ~~~~~ ~~~ +
  2 \Omega_z
  =
  \Delta_{1,xy}({\mathcal{N}}).
  \label{Eq1-Omega1}
\end{eqnarray}
The solution of these equations is
\begin{eqnarray}
  B_z &=&
  \frac{2 I + 1}{g \mu_B}~
  \frac{\hbar
        \big[\Delta_{1,xy}(10) - \Delta_{1,xy}(5)\big]}
       {\tilde\beta(10) - \tilde\beta(5)},
  \label{Bz-res}
  \\
  \Omega_z &=&
  \frac{\tilde\beta(10)~\Delta_{1,xy}(5) -
        \tilde\beta(5)~\Delta_{1,xy}(10)}
       {\tilde\beta(10) - \tilde\beta(5)},
  \label{Omega-z-res}
\end{eqnarray}
where
\begin{eqnarray*}
  \tilde\beta({\mathcal{N}}) &=&
  5 \beta_{0,5/2}^{z}({\mathcal{N}})) -
  3 \beta_{0,3/2}^{z}({\mathcal{N}})).
\end{eqnarray*}

With the optical lattice arranged in $y$-$z$ and $z$-$x$ planes, measurements can 
be made of $\Delta_{1,yz}({\mathcal{N}})$ and
$\Delta_{1,zx}({\mathcal{N}})$ for ${\mathcal{N}}=5,10$.
These measurements allow us to find $B_x$, $B_y$,
$\Omega_x$ and $\Omega_y$,
\begin{eqnarray}
  B_x &=&
  \frac{2 I + 1}{g \mu_B}~
  \frac{\hbar
        \big[\Delta_{1,yz}(10) - \Delta_{1,yz}(5)\big]}
       {\tilde\beta(10) - \tilde\beta(5)},
  \label{Bx-res}
  \\
  B_y &=&
  \frac{2 I + 1}{g \mu_B}~
  \frac{\hbar
        \big[\Delta_{1,zx}(10) - \Delta_{1,zx}(5)\big]}
       {\tilde\beta(10) - \tilde\beta(5)},
  \label{By-res}
\end{eqnarray}
\begin{eqnarray}
  \Omega_x &=&
  \frac{\tilde\beta(10)~\Delta_{1,yz}(5) -
        \tilde\beta(5)~\Delta_{1,yz}(10)}
       {\tilde\beta(10) - \tilde\beta(5)},
  \label{Omega-x-res}
  \\
  \Omega_y &=&
  \frac{\tilde\beta(10)~\Delta_{1,yz}(5) -
        \tilde\beta(5)~\Delta_{1,yz}(10)}
       {\tilde\beta(10) - \tilde\beta(5)}.
  \label{Omega-y-res}
\end{eqnarray}

With the optical lattice arranged in the $x$-$y$, $y$-$z$ and $z$-$x$ planes, 
measurements can be made of
$\Delta_{0,xy}$, $\Delta_{0,yz}$ and $\Delta_{0,zx}$
for ${\mathcal{N}}=10$. Three equations are thereby obtained,
\begin{eqnarray}
  &&
  \bigg(
       \frac{g \mu_B \beta_{0,1/2}^{\parallel}}{2 I+1}~B_{\alpha} +
       M \varrho_{0,1/2}^{\parallel} a_{\alpha}
  \bigg)^{2} +
  \nonumber \\ &&
  \bigg(
       \frac{g \mu_B \beta_{0,1/2}^{\parallel}}{2 I+1}~B_{\alpha'} +
       M \varrho_{0,1/2}^{\parallel} a_{\alpha'}
  \bigg)^{2} +
  \nonumber \\ &&
  \bigg(
       \frac{g \mu_B \beta_{0,1/2}^{z}}{2 I+1}~
       B_{\alpha''} +
       \hbar \Omega_{\alpha''}
  \bigg)^{2}
  ~=~
  \hbar^2 \Delta_{0,\alpha \alpha'}^{2},
  \label{Eq2-Omega0}
\end{eqnarray}
where $(\alpha,\alpha',\alpha'') = (x,y,z)$, $(y,z,x)$ and $(z,x,y)$.
Here $B_{\alpha}$ are given by Eqs.~(\ref{Bz-res}), (\ref{Bx-res})
and (\ref{By-res}) for $\alpha=z,x,y$, whereas
$\Omega_{\alpha}$ are given by Eqs.~(\ref{Omega-z-res}),
(\ref{Omega-x-res}) and (\ref{Omega-y-res}).
The solution of Eq.~(\ref{Eq2-Omega0}) is
\begin{eqnarray}
  a_x &=&
  \frac{1}{M \varrho_{0,1/2}^{\parallel}}~
  \Bigg\{
       \frac{1}{2}~
       \sqrt{A_{x,y}+A_{z,x}-A_{y,z}}
  \nonumber \\ && -
       \frac{g \mu_B \beta_{0,1/2}^{\parallel}}{2 I+1}~B_x
  \Bigg\},
  \label{a-x-res}
  \\
  a_y &=&
  \frac{1}{M \varrho_{0,1/2}^{\parallel}}~
  \Bigg\{
       \frac{1}{2}~
       \sqrt{A_{y,z}+A_{x,y}-A_{z,x}}
  \nonumber \\ && -
       \frac{g \mu_B \beta_{0,1/2}^{\parallel}}{2 I+1}~B_y
  \Bigg\},
  \label{a-y-res}
  \\
  a_z &=&
  \frac{1}{M \varrho_{0,1/2}^{\parallel}}~
  \Bigg\{
       \frac{1}{2}~
       \sqrt{A_{z,x} + A_{y,z} - A_{x,y}}
  \nonumber \\ && -
       \frac{g \mu_B \beta_{0,1/2}^{\parallel}}{2 I+1}~B_z
  \Bigg\},
  \label{a-z-res}
\end{eqnarray}
where
\begin{eqnarray*}
  A_{\alpha,\alpha'} &=&
  \hbar^2 \Delta_{0,\alpha \alpha'}^{2} -
  \bigg(
       \frac{g \mu_B \beta_{0,1/2}^{z}}{2 I+1}~
       B_{\alpha''} +
       \hbar \Omega_{\alpha''}
  \bigg)^{2},
\end{eqnarray*}
and $(\alpha,\alpha',\alpha'') = (x,y,z)$, $(y,z,x)$ and $(z,x,y)$.

As already discussed in Sec.~\ref{SubSec:Raman} in connection with the far-off-resonance 
Raman transition $|0,1/2\rangle \leftrightarrow |0,-1/2\rangle$, to determine the Raman 
resonance condition $\omega_p - \omega_s = \Delta$ in {\em all} the far-off-resonance Raman 
processes considered in this section ($|0,1/2\rangle \leftrightarrow |0,-1/2\rangle$,
$|0,3/2\rangle \leftrightarrow |0,5/2\rangle$ and $|0,-3/2\rangle \leftrightarrow 
|0,-5/2\rangle$), one can employ the Ramsey time-separated oscillating field method 
\cite{Ramsey_50} with Raman pulses \cite{Zanon}.

\clearpage



\begin{thebibliography}{99}
\bibitem{other-symmetry}
It is also of interest to explore SDOLP with other symmetries.
\bibitem{band_structure}
One could do a full band-structure calculation, but 
our interest is in the lowest energy states with 
negligible well to well tunneling and wave functions localized 
at small $r$, hence the 2D isotropic approximation
is sufficient.
\end{thebibliography}
\end{document}